\documentclass{aastex}
\usepackage{emulateapj5}
\usepackage{apjfonts}
\usepackage{epsf}

\slugcomment{IPMU09-0111}

\shorttitle{Does GMF Disturb the Correlation of HECRs with their Sources?}
\shortauthors{Takami \& Sato}

\begin{document}

\title{Does Galactic Magnetic Field Disturb the Correlation of the Highest Energy Cosmic Rays with their Sources?}

\author{Hajime Takami\altaffilmark{1} and Katsuhiko Sato\altaffilmark{1,2}}
\email{hajime.takami@ipmu.jp}

\altaffiltext{1}{Institute for the Physics and Mathematics of the Universe, the University of Tokyo, 5-1-5, Kashiwanoha, Kashiwa, Chiba 277-8583, Japan}
\altaffiltext{2}{Department of Physics, School of Science and Engineering, Meisei University, 2-1-1, Hodokubo, Hino-shi, Tokyo 191-8506, Japan}

\begin{abstract}
The propagation trajectories of the highest energy cosmic rays (HECRs) are deflected by not only intergalactic magnetic field but also Galactic magnetic field (GMF). These magnetic fields can weaken the positive correlation between the arrival directions of HECRs and the positions of their sources. In order to explore the effect of GMF on the expected correlation, we simulate the arrival distribution of protons with the energy above $6 \times 10^{19}$ eV taking several GMF models into account, and then test the correlation between the protons and their sources assumed in the simulation. The dependence of the correlation signals on GMF models are also investigated. The correlation can be observed by accumulating $\sim 200$ protons in a half hemisphere. Typical angular scale at which the positive signal of the correlation is maximized depends on the spiral component of GMF models. That angular scale is $\sim 5^{\circ}$ for bisymmetric spiral (BS) GMF models and $\sim 7^{\circ}$ for axisymmetric spiral (AS) GMF models if the number density of HECR sources, $n_s$, is $\sim 10^{-4}$ Mpc$^{-3}$. An additional vertical (dipole) component of GMF affects these angular scale by $0.5^{\circ}$ - $1^{\circ}$. The difference between the correlation signal for the BS models and that for the AS models is prominent in the northern sky. Significance of the positive correlation depends on source distribution. The probability that the number of simulated HECR events correlating with sources is smaller than the number of random events correlating with the same sources by chance is much less than $10^{-3}$ ($\sim 3\sigma$) in almost all the source distributions with $n_s = 10^{-4}$ Mpc$^{-3}$ under 200 protons detection, but $\sim 10\%$ of source distributions predicts the chance probability more than $10^{-3}$ in the AS GMF model. In addition, we also briefly discuss the effect of GMF for heavy-nuclei dominated composition. 
\end{abstract}

\keywords{cosmic rays --- methods: numerical --- ISM: magnetic fields --- 
Galaxy: structure}

\section{Introduction} \label{introduction}

The origin of the highest energy cosmic rays (HECRs) is one of the biggest mysteries in astrophysics. Although theoretical candidates of HECR sources have been proposed (see \cite{bhattacharjee00,torres04} for reviews), HECR sources have not been successfully identified. A difficulty to identify HECR sources is the existence of Galactic magnetic field (GMF) and intergalactic magnetic field (IGMF). Since HECRs are charged, these magnetic fields deflect the trajectories of HECRs from their sources to the Earth, i.e., the arrival directions of HECRs are shifted from the positions of their sources.

Recent progress to construct HECR observatories with larger effective area has enabled us to accumulate a number of HECR events and to draw anisotropy in the arrival distribution of HECRs \citep{takeda99,abraham07,abraham08}. The anisotropy includes information on the positions of their sources and intervening magnetic fields from the sources to the Earth. Also, the Pierre Auger Observatory (PAO) found the correlation between the arrival directions of 27 HECR events above $5.7 \times 10^{19}$ eV and the positions of extragalactic astrophysical objects in local Universe within $z = 0.018$ with angular scale of $3.1^{\circ}$ \citep{abraham07,abraham08}. Several authors have also analyzed the PAO events. \cite{george08} showed significant correlation of the PAO events with hard X-ray selected active galactic nuclei (AGNs) by using the 2 dimensional generalization of Kolmogorov-Smilnov test. Generally, the correlation between the PAO events and galaxies, which are a representative of matter distribution in local Universe, has been shown \citep{ghisellini08,kashti08,takami09c}. \cite{takami09c} estimated the typical angular scale of positive correlation between HECRs and galaxy distribution and bounded the deflection angles of HECRs in local Universe. All these works indicate that the origin of HECRs is extragalactic objects distributed with being comparable to galaxies. Furthermore, in an optimistic view, we can interpret these results as the evidence that intervening magnetic fields are not so strong that the arrival directions of HECRs lose information on the positions of their sources as predicted by \cite{dolag05} and \cite{takami06} (but see \cite{sigl04}, \cite{kotera08}, \cite{das08}, and \cite{ryu09} for strong IGMF). This interpretation implies that a more number of HECR events can unveil the positions of their sources as strong event clusters within small angular scale, as demonstrated by \cite{blasi04} without magnetic field. \cite{takami08a} showed that such a signal is also predicted for the highest energy protons (HEPs) even considering a structured IGMF model, which reflects the matter distribution of local Universe actually observed. We can regard these works as predictions of the PAO result. However, it is still under debate that astrophysical objects which correlate with the PAO events are really HECR sources because these objects have weaker activity than theoretically predicted source candidates \citep*{moskalenko09}.

However, a recent PAO result showed that the significance of the correlation reported in \cite{abraham07,abraham08} decreased, though the correlation is still significant at more than 2$\sigma$ \citep{hague09}. In addition, the High Resolution Fly's Eye (HiRes) reported that their is no correlation between HiRes events and extragalactic objects based on a similar analysis to the analysis of the PAO \citep*{abbasi08}. Thus, we should reconsider the correlation between HECRs and their sources, and therefore the effect of intervening magnetic fields. The large deflection angles of HECRs could weaken the correlation. The possibilities that the deflection of HECR trajectories is larger than one expected are (i) GMF affects the deflections significantly (ii) the main composition of the HECRs is heavier than protons (iii) the IGMF model used in \cite{takami08a} was optimistic, i.e., the deflection of HEPs in this IGMF model is smaller than that in the real Universe. In this study, we focus on the possibility (i) assuming that the composition of HECRs are purely protons and IGMF subdominantly affects the deflections of HEPs. The other possibilities will be discussed in \S \ref{conclusion}.

Cosmic rays arriving at the Earth are inevitably affected by GMF. This importance has driven us to consider the propagation of HECRs in the Galactic space \citep*{stanev97,medinatanco98,harari99,kalashev01,alvarez02,harari02a,harari02b,tinyakov02,prouza03,yoshiguchi03,yoshiguchi04,tinyakov05,kachelriess07,takami08b,cuoco08b,golup09}. Many of these works have mainly investigated the propagation of HECRs itself in detail. Based on the results of these studies, the deflection angles of the trajectories of HEPs are highly dependent on their arrival directions reflecting the structure of GMF and can be less than $\sim 10^{\circ}$ at the highest energy ($\sim 10^{20}$ eV) except in the direction of the Galactic center. The angular dependence of the deflection of HEPs distorts the apertures of HECR detectors to extragalactic space, so-called magnetic lensing effect \citep*{harari99,harari02a,harari02b,kachelriess07}. Some of them guessed HECR sources using experimental data and HECR trajectories calculated in GMF models. \cite{golup09} studied the possibility to reconstruct the positions of HECR sources and several properties of intervening magnetic field from detected HECRs. \cite{alvarez02} and \cite{yoshiguchi03} calculated the arrival distribution of HECRs assuming source models, and then analyzed the arrival distribution. However, they mainly investigated the auto-correlation of HECRs. The cross-correlation between HECRs and their sources was hardly considered.

In this study, we investigate the cross-correlation between the arrival directions of HEPs and the positions of their sources containing the effect of GMF. For this purpose, we simulate the arrival distribution of protons above $6 \times 10^{19}$ eV taking their propagation in magnetized Galactic space into account. Since the positions of sources of simulated HEPs are known in simulations, we can calculate cross-correlation functions between HECRs and their sources. The modelling of GMF is still controversial, as different models of GMF have been adopted in the previous studies introduced above. Thus, we also examine the dependence of predicted correlation signals on GMF models.

This paper is laid out as follows: \S \ref{calculation} is devoted to explain GMF models, a HEP source model, the calculation method of the arrival distribution of HEPs, and a statistical quantity to use our correlation analysis. \S \ref{results} is dedicated to giving calculation results. First of all, we investigate the effect of GMF to the propagation trajectories of HEPs in \S \ref{def}. Then, we calculate the cross-correlation between the arrival directions of simulated HEPs and their nearby sources, and examine the angular scale at which positive correlation is predicted in \S \ref{ccor_sources}. In \S \ref{ccor_best}, we estimate the angular distance within which positive correlation signals are maximized and its significance. Finally, we summarize results obtained in \S \ref{results} and briefly discuss the possible effect of IGMF and heavy-nuclei dominated composition.

\section{Calculation Tools} \label{calculation}

\subsection{GMF Models} \label{gmf}

GMF can be divided into 3 components; a spiral component following the optical spiral arm of the Galaxy, a vertical component, and a turbulent component. These components of GMF are modelled as follows. 

The spiral component could be well described by axisymmetric (AS) or bisymmetric (BS) field models. BS models have several field reversals, while AS models do not. Although a field reversal at $\sim 0.5$ kpc inside from the solar system is well established, the existence of the other field reversals predicted by BS models, i.e., which is a better modelling, is still controversial. Recognizing this uncertainty, we adopt both BS and AS models in this study and discuss the dependence of the positional correlation between HECRs and their sources on GMF models. We use the parametrization of both GMF models by \cite{stanev97}, described below.

The radial and azimuthal components of a spiral magnetic field in the Galactic plane are given by 
\begin{equation}
B_{r_{||}} = B(r_{||}, \theta_{||}) \sin p, ~~
B_{\theta} = B(r_{||}, \theta_{||}) \cos p, 
\end{equation}
where $r_{||}$ and $\theta_{||}$ are the galactocentric distance and azimuthal angle around the Galactic center, respectively. Note that $\theta_{||}$ is defined as increasing clockwise and $\theta_{||} = 0^{\circ}$ corresponds to the direction of the Galactic center. $p$ is the pitch angle of the spiral field in the neighborhood of the solar system, set to $p = -10^{\circ}$. The field strength at a point $(r_{||}, \theta_{||})$ in the Galactic plane is written as 
\begin{equation}
B(r_{||},\theta_{||}) = \left\{
\begin{array}{ll}
b(r_{||}) \cos \left( \theta_{||} - \beta \ln \frac{r_{||}}{r_0} \right) & :~{\rm BS} \\
b(r_{||}) \left| \cos \left( \theta_{||} - \beta \ln \frac{r_{||}}{r_0} \right) \right| & :~{\rm AS}.
\end{array}
\right. 
\end{equation}
Here $\beta \equiv \left( \tan p \right)^{-1} = -5.67$ and $r_0 = 10.55$ kpc is the galactocentric distance of the location with maximum field strength at $\theta_{||} = 0^{\circ}$, which can be expressed as $r_0 = (R_{\oplus} + d) \exp \left[ -(\pi/2) \tan p \right]$, where $R_{\oplus} = 8.5$ kpc is the distance of the solar system from the Galactic center and $d = -0.5$ kpc is the distance to the nearest field reversal from the solar system in the BS model. Negative $d$ means that the nearest field reversal takes place in the direction of the Galactic center. Note that the AS model does not have any field reversals. $b(r_{||})$ is the radial profile of the magnetic field strength modelled by 
\begin{equation}
b(r_{||}) = B_0 \frac{R_{\oplus}}{r_{||}}, 
\end{equation}
where $B_0 = 4.4 \mu$G, which corresponds to $1.5 \mu$G in the neighborhood of the solar system. In the region around the Galactic center ($r_{||} < 4$ kpc), the field is highly uncertain and therefore assumed to be constant and equal to its value at $r_{||} = 4$ kpc. The spiral field is assumed to be zero for $r_{||} > 20$ kpc.

For the spiral halo field, we adopt an exponential decreasing model with two scale heights \citep*{alvarez02}, 
\begin{equation}
B(r_{||},\theta_{||},|z|) = B(r_{||},\theta_{||}) 
\left\{
\begin{array}{ll}
\exp(-|z|) & :~|z| \leq 0.5~{\rm kpc} \\ 
\exp(\frac{-|z|}{4}-\frac{3}{8}) & :~|z|>0.5~{\rm kpc} 
\end{array}
\right.
\end{equation} 
where the factor $\exp(-3/8)$ makes the field continuous on $z$. The parity is represented as 
\begin{equation}
B(r_{||},\theta_{||},-z) = \left\{
\begin{array}{ll}
B(r_{||},\theta_{||},z) & :~{\rm S~type~parity} \\
-B(r_{||},\theta_{||},z) & :~{\rm A~type~parity}. 
\end{array}
\right.
\end{equation}
The combination between the 2 spiral structures and the parity produces 4 different spiral GMF models.

There is little observational evidence of the vertical component of GMF. Near the solar system, a vertical component of magnetic field with strength of 0.2 - 0.3 $\mu$G, which is directed toward the north Galactic pole, is observed \citep*{han94}. Also, a lot of nonthermal gaseous filaments perpendicular to the Galactic plane with tens of $\mu$G to mG have been discovered near the Galactic center \citep*{han07}. Little information allows us to construct the realistic model of the vertical component of GMF. \cite{stanev97} assumed uniform magnetic field with $z$-direction which is a possible effect of the existence of Galactic wind. Here, we adopt a dipole field as a vertical component of GMF following \cite{alvarez02}. A dipole field is described as 
\begin{eqnarray}
\begin{array}{l}
B_x = -3 \mu_{\rm G} \sin \phi \cos \phi \cos \varphi / r^3, \\
B_y = -3 \mu_{\rm G} \sin \phi \cos \phi \sin \varphi / r^3, \\
B_z = \mu_{\rm G} ( 1 - \cos \phi ) / r^3, 
\end{array}
\end{eqnarray}
where $\phi$ and $\varphi$ are the zenith angle and the azimuthal angle in spherical coordinates centered at the Galactic center, respectively. $\mu_{\rm G} = 184.2 \mu$G kpc$^3$ is the magnetic moment of the Galactic dipole, which is normalized at 0.3$\mu$G in the vicinity of the solar system. Theoretically, a dipole field can be predicted by the so-called A0 mode in the dynamo theory \citep*{widrow02,beck09}. Since the A0 mode predicts the A-type parity of a spiral component, it is non-trivial that GMF models with S-type parity have a dipole component naturally. Note that there is no clear observational evidence of a dipole magnetic field.

GMF has also a significant turbulent component, but its field strength, $B_{\rm turb}$, has large uncertainty. Several observations have estimated as $B_{\rm turb} = $(0.5 - 2) $B_{\rm reg}$ \citep*{beck01}, where $B_{\rm reg}$ is the strength of the regular component of GMF. In this study, we assume $B_{\rm turb} = B_{\rm reg}$ and $B_{\rm reg}$ is the sum of the strength of the spiral component and the strength of the dipole component. The coherent length of the turbulent component, $l_c$, has been thought to be typically $\leq 100$ pc \citep*{beck01}. We assume a turbulent magnetic field with the Kolmogorov power spectrum with $l_c = 50$ pc. Since this $l_c$ is much smaller than the Larmor radius of HEPs the turbulent component does not significantly affect the propagation of HEPs with $\sim 10^{20}$ eV \citep*{tinyakov05}. However, this component will be important for the propagation of heavy nuclei in the Galactic space due to their large charges. 

In this study, we adopt a turbulent IGMF model constructed 
in \cite{yoshiguchi03b}. 
We cover Galactic space with cubes of side $l_c$ 
and assume that turbulent magnetic field in each cube is represented 
as a Gaussian random field with zero mean and a power-low spectrum 
\begin{eqnarray}
\left< B({\bf k}) B^*({\bf k}) \right> \left\{
\begin{array}{ll}
\propto k^{n_{\rm H}} & {\rm for}~~2\pi/l_c \leq k \leq 2\pi/l_{\rm cut}, \\
= 0 & {\rm otherwise}, 
\end{array}
\right.
\end{eqnarray}
where $l_{\rm cut}$ is a numerical cutoff scale. 
Physically, one expects $l_{\rm cut} \ll l_c$, 
but we set $l_{\rm cut} = \frac{1}{8} l_c$ to save the CPU time. 
We adopt $n_{\rm H} = -11/3$, which corresponds to the Kolmogorov 
power spectrum. 

\begin{figure*}
\epsscale{0.85}
\plotone{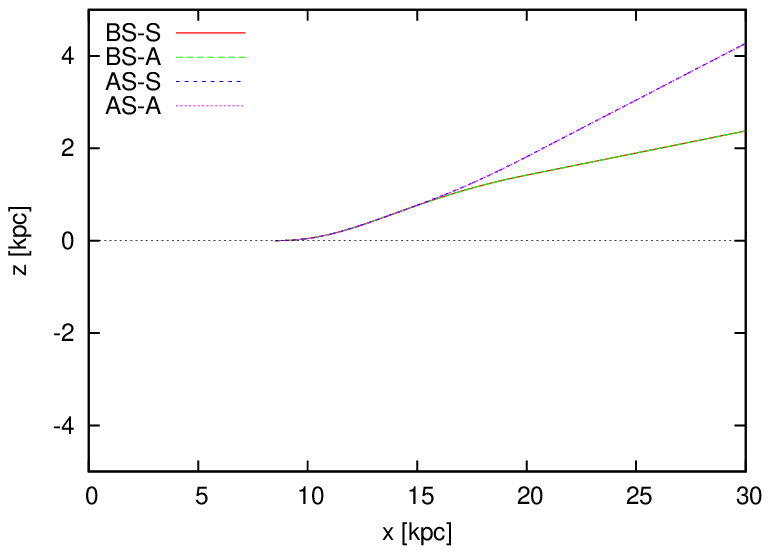} \\ 
\plotone{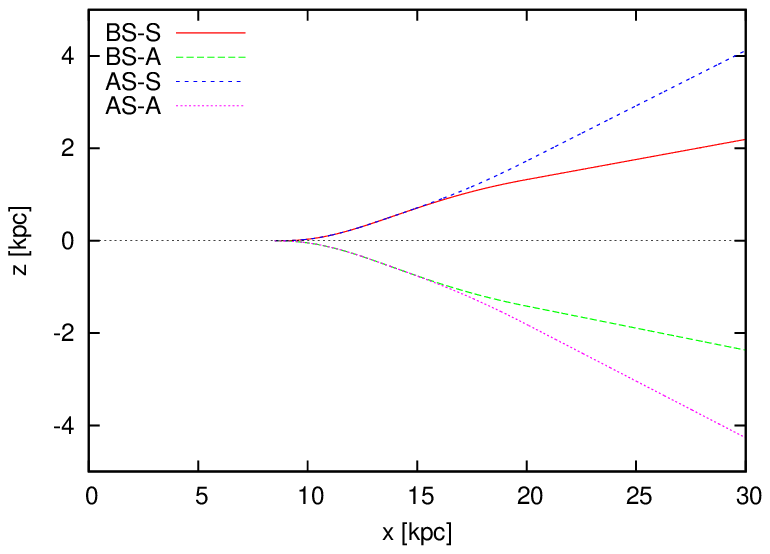}
\plotone{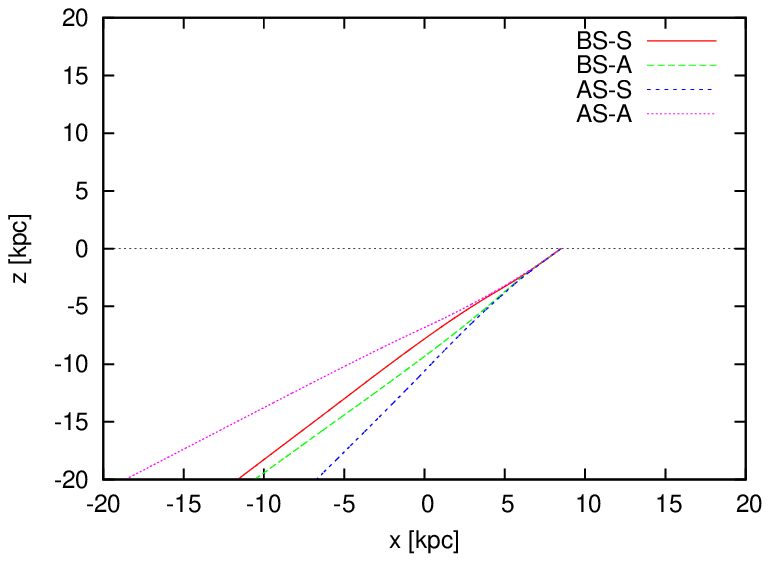}
\caption{Examples of trajectories of protons with the energy of $10^{19.8} = 6.3 \times 10^{19}$ eV projected onto a plane perpendicular to the Galactic disk and including the Galactic center and the Earth. The position of the Earth is $x = 8.5$ kpc. Only the spiral component of GMF is considered. The arrival directions of protons are $(\ell, b) = (180.0^{\circ}, 0.3^{\circ})$ ({\it upper panel}), $(180.0^{\circ}, -0.3^{\circ})$ ({\it lower left panel}), and $(0.0^{\circ}, -44.8^{\circ})$ ({\it lower right panel}). The 4 trajectories in each panel correspond to the trajectories of protons in the BS-S model ({\it red}), BS-A model ({\it green}), AS-S model ({\it blue}), and AS-A model ({\it magenta}).
\label{fig:trajectories}}
\end{figure*}

\subsection{Source model of HEPs} \label{sources}

The PAO data implies that HECR sources are distributed to be comparable with matter (or galaxy) distribution in local Universe \citep*{abraham07,abraham08,ghisellini08,kashti08,takami09c}. Following this result, we consider HEP source model in which sources are distributed following galaxy distribution actually observed \cite{takami06}. This model is summarized as below. See \citep*{takami06} for more details.

This source model is based on the Infrared Astronomical Satellite Point Source Redshift Survey (IRAS PSCz) catalog of galaxies \citep{saunders00}. The IRAS PSCz catalog is appropriate to construct the model of HEP sources distributed with being comparable to galaxy distribution because this galaxy catalog is a flux-limited, i.e., uniform, catalog of galaxies with very large sky coverage ($\sim$ 84\% on the whole sky). This catalog includes 2 selection effects: One is the effect of undetected dark galaxies with the flux below the flux-limit (0.6 Jy), and the other is $\sim 16\%$ of unobserved area which mainly corresponds to the sky near the Galactic plane. We correct these selection effects by using a luminosity function of the IRAS galaxies \citep{takeuchi03}. The former effect can be corrected by adding virtual galaxies with the flux less than the flux-limit in the vicinity of the original IRAS galaxies not to spoil the original structured galaxy distribution. For the latter effect, we simply assume that galaxies are distributed isotropically in the unobserved area. These corrections allow us to construct (a model of) a complete galaxy catalog. We adopt this corrected galaxy catalog only within 200 Mpc to make HECR source distribution because the number of original galaxies outside 200 Mpc is too small to make source distribution reflecting matter distribution in the Universe. It is assumed that galaxies are isotropically distributed outside 200 Mpc up to 1 Gpc.

The subsets of this corrected galaxy catalog are regarded as HEP source distributions, i.e., we randomly select galaxies from this corrected catalog according to a given number density of HEP sources, $n_s$, and then the selected galaxies are regarded as HEP sources. The distribution of HEP sources traces the distribution of the original IRAS galaxies within 200 Mpc. HEP sources outside 200 Mpc are distributed isotropically by definition, but these isotropic sources hardly contribute to the total flux of HEPs because more than 90\% of HEPs above $6 \times 10^{19}$ eV detected at the Earth is generated within 200 Mpc.

For the physical properties of HEP sources, we assume that all the sources emit HEPs persistently with the same power and with a power-law spectrum of $\propto E^{-2.6}$, where $E$ is the energy of HEPs. This spectral index can well reproduce the observed spectrum of HECRs above $10^{19}$ eV. We consider 2 number densities of HEP sources, $n_s = 10^{-4}$ and $10^{-5}$ Mpc$^{-3}$. $10^{-4}$ Mpc$^{-3}$ can best reproduce anisotropy in the arrival distribution of HECRs detected by the PAO \citep*{takami09a,cuoco08}, while $10^{-5}$ Mpc$^{-3}$ is the number density to reproduce the AGASA anisotropy well \citep*{takami06,blasi04,kachelriess05}.

When selecting HEP sources from the corrected galaxy catalog, we do not consider any specific condition, e.g., selecting only AGNs and so on. Thus, we generate 500 source distributions for $n_s = 10^{-4}$ Mpc$^{-3}$ and $n_s = 10^{-5}$ Mpc$^{-3}$, and discuss the cross-correlation between HEPs and their sources statistically.

\subsection{Simulation of Arriving HEPs} \label{simulation}

In order to simulate the arrival distribution of HEPs for a given source distribution, we adopt a calculation method developed by \cite{takami06}. When a cosmic ray is injected into magnetized Universe from its source, whether this cosmic ray can reach the Earth or not cannot be known until its trajectory is actually calculated because of the existence of magnetic field. Such a cosmic ray does not arrive at the Earth in many case. This requires the injection of many particles and the calculation of their trajectories to simulate the arrival distribution of HECRs at the Earth. It is time-wasting to calculate the trajectories of HECRs which cannot reach the Earth. Our method solved this problem by considering only the trajectories of HEPs reaching the Earth, and enables us to calculate the arrival distribution of protons within reasonable CPU time.

In this method, the trajectories of protons are calculated by the backtracking method. The trajectory of an oppositely-charged proton ejected from the Earth can regarded as that of a proton coming from extragalactic space. Many oppositely-charged protons are ejected from the Earth isotropically, and their trajectories in the Galactic space are calculated. A decade of the energy of HEPs is divided into 10 bins in logarithmic scale. 125,000 particles are emitted from the Earth per energy bin in this study. We consider HEPs with the energy from $10^{19}$ eV to $10^{21}$ eV at the Earth. Once the ejected particles escape from the Galactic space, they propagate straightforwardly in intergalactic space with energy-{\it gain}. The boundary between the Galactic and extragalactic space is set to be 40 kpc from the Galactic center. The calculation of their trajectories is finished when the comoving distance of the particles exceeds 1,500 Mpc or the energy of the particles exceeds $10^{22}$ eV. This energy corresponds to the maximum acceleration energy of protons at sources. As the energy-loss processes of HEPs (the energy-{\it gain} processes of oppositely-charged protons), in addition to Bethe-Heitler pair creation and photopion production by interactions with cosmic microwave background (CMB) photons, adiabatic energy-loss due to the cosmic expansion is also considered in intergalactic space. All the energy-loss can be neglected during propagation in the Galactic space because all the energy-loss lengths of protons are much larger than the scale of the Galactic space. We assume the standard $\Lambda$CDM cosmology with $\Omega_m = 0.3$, $\Omega_{\Lambda} = 0.7$, $H_0 = 71$ km s$^{-1}$ Mpc$^{-1}$. An analytical fitting formula of the energy-loss length of Bethe-Heitler pair creation by \cite{chodorowski92} is used. For photopion production, we adopt inelasticity and mean-free path calculated by an event generator SOPHIA \citep*{mucke00}.

\begin{figure*}
\epsscale{1.90}
\plotone{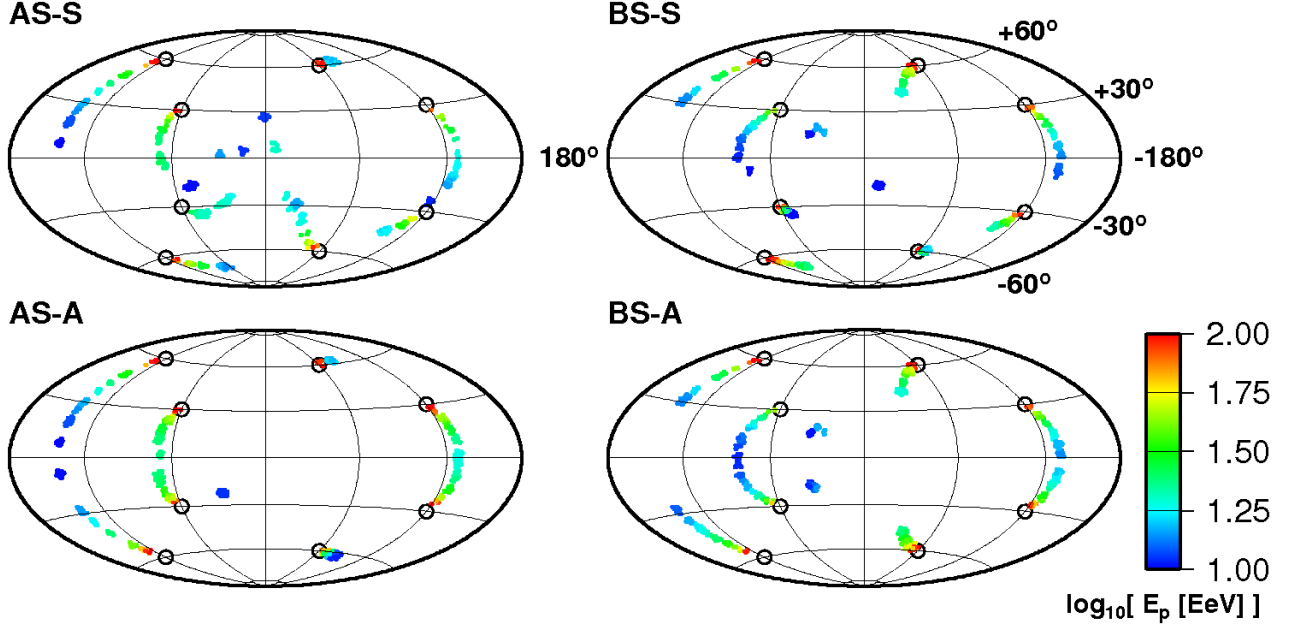}
\caption{Arrival directions of protons with the energy above $10^{19}$ eV emitted from 8 sources located at 50 Mpc from the Earth ({\it black circles}) in Galactic coordinates. The energy of each proton is represented in color. The 4 different spiral components of GMF are taken into account.
\label{fig:ad}}
\end{figure*}

When the trajectory of the $i$th oppositely-charged proton passes over sources, a weight below is attached into this trajectory for a positive arriving proton, 
\begin{equation}
P_{\rm selec}(E,i) \propto \sum_j \frac{1}{( 1 + z_{i,j} ) {d_{i,j}}^2} 
\frac{dN/dE_g ({E_g}^j)}{E^{-1.0}} \frac{dE_g}{dE}, 
\end{equation}
where $j$ labels sources on the trajectory. $z_{i,j}$ and $d_{i,j}$ are the redshift and comoving distance of the $j$th source, respectively. ${E_g}^j$ is the energy of the particle at the $j$th source. $dN/dE_g (E_g) \propto {E_g}^{-2.6}$ is the injection spectrum of HEPs at the $j$th source. $dE_g/dE$ can be calculated by the propagation data. $E^{-1.0}$ is a factor which originates from the ejection spectrum of oppositely-charged protons at the Earth ($dN/d\log_{10}E \propto$ const.). This weight corresponds to a relative probability in order that the $i$th particle is a really arriving proton. This factor is attached to all the trajectories of oppositely-charged protons. In order to obtain protons arriving at the Earth, we randomly select the required number of trajectories according to the probability proportional to $P_{\rm selec}(E,i)$. The energy and arrival directions of the protons are the energy and the injection directions of the corresponding oppositely-charged protons. We assume that the arrival directions fluctuates following a 2 dimensional Gaussian distribution with $\sigma = 1^{\circ}$ to take the finite accuracy of the arrival direction determination of HECR observatories into account.

\subsection{Statistical Methods} \label{ccor}

We will discuss the correlation between the arrival directions of HEPs and their sources in the next section. A cross-correlation function for this purpose is defined as \citep{takami09c,blake06}, 
\begin{equation}
w_{es}(\theta) = \frac{ES(\theta) - ES'(\theta) - E'S(\theta) + E'S'(\theta)}{E'S'(\theta)}, 
\end{equation}
where $E$ and $S$ represent HEP events and their sources which are considered to calculate the cross-correlation function, respectively. $ES(\theta)$ is the number of pairs between $E$ and $S$ with the separation angles from $\theta$ to $\theta + \Delta \theta$ divided by $N N_s$ for normalization, where $N$ and $N_s$ are the number of events and the number of sources considered, respectively. $\Delta \theta$ is set to be $1^{\circ}$. $E'$ and $S'$ represent randomly distributed events following the aperture of a HECR observatory and randomly distributed sources following the selection window of a source catalog, respectively. Here, $S'$ represents just sources randomly distributed in the sky because the selection effects were fixed. $ES'(\theta)$, $E'S(\theta)$, and $E'S'(\theta)$ are defined following the definition of $ES(\theta)$. $E'$ and $S'$ allow to correct their non-uniform apertures. By definition, $w_{es}(\theta) > 0$ means positive correlation.

The aperture of a ground array is simply described as \citep{sommers01} 
\begin{equation}
\omega(\delta) \propto \cos (a_0) \cos (\delta) \sin (\alpha_m) 
+ \alpha_m \sin (a_0) \sin (\delta), 
\end{equation}
where $\alpha_m$ is given by 
\begin{equation}
\alpha_m = \left\{ 
\begin{array}{ll}
0 & {\rm if}~\xi > 1 \\
\pi & {\rm if}~\xi < -1 \\
\cos^{-1}(\xi) & {\rm otherwise}
\end{array}
\right.
\end{equation}
and 
\begin{equation}
\xi \equiv \frac{\cos (\Theta) - \sin (a_0) \sin (\delta)}
{\cos (a_0) \cos (\delta)}. 
\end{equation}
Here, $a_0$ and $\Theta$ are the terrestrial latitude of a ground array and the zenith angle for an experimental cut. We simulate the apertures of 2 HECR observatories; the PAO and Telescope Array (TA). These 2 observatories are complementary because the PAO observes the southern sky while the TA observes the northern sky. Thus, we consider these 2 apertures as representatives of HECR observatories in the southern and northern sky. Parameters are $a_0 = -35.2^{\circ}$ and $\Theta = 60^{\circ}$ for the PAO \citep{abraham08}, and $a_0 = 39.3^{\circ}$ and $\Theta = 45^{\circ}$ for the TA \citep{nonaka09}.

\begin{figure*}
\epsscale{1.90}
\plotone{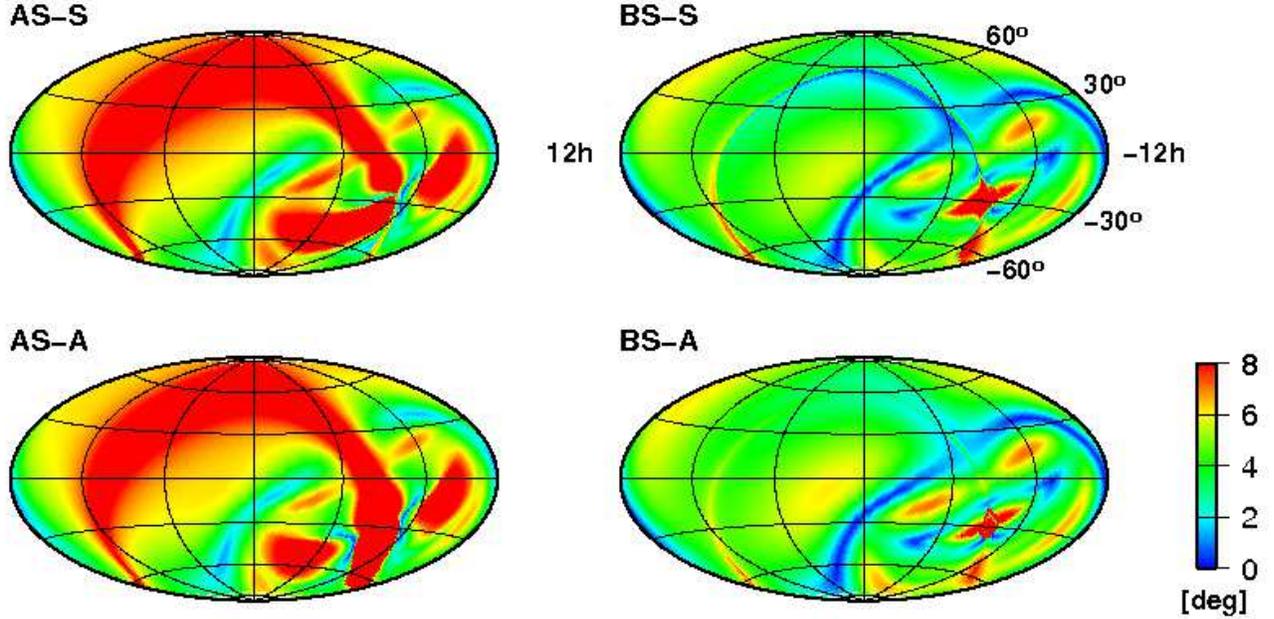}
\caption{Distribution of the deflection angles of protons with the energy of $10^{19.8}$ eV by the spiral component of GMF as a function of the arrival directions of them in equatorial coordinates. Their deflection angles are shown in color. The difference between the AS models and BS GMF models is prominent especially in the northern hemisphere.} 
\label{fig:def}
\end{figure*}

\section{Results} \label{results}

\subsection{Deflections of protons by GMF} \label{def}

Before considering the correlation between the arrival directions of HEPs and their sources, we examine the trajectories of HEPs in GMF.

The dipole component of GMF is directed toward the north Galactic pole in the Galactic plane, and therefore protons coming from extragalactic space propagate clockwise, as viewed from the north Galactic pole. This fact indicates the arrival directions of HEPs are shifted from the positions of their sources to lower Galactic longitude. This effect is weaker in the direction of the Galactic anti-center because of the smaller field strength than in the direction of the Galactic center. The strong dipole field near the Galactic center makes the complex trajectories of HEPs. In addition, since the dipole field above and below the Galactic center has the opposite direction to that in the Galactic plane away from the Galactic center, the trajectories of HEPs from the direction of $\ell \sim 0^{\circ}$, where $\ell$ is the Galactic longitude, can be deflected anticlockwise in the direction of $\ell \sim 0^{\circ}$ \citep{yoshiguchi04,takami08b}. The deflection angles of HEPs are not so large that the correlation disappears completely ($< 10^{\circ}$) except the direction of the Galactic center if protons above $10^{19.8}$ eV are considered \citep{takami08b}.

The turbulent component of GMF makes HEPs propagate diffusively. The typical deflection angle of a HEP is 
\begin{equation}
\theta (\epsilon_p, D_{\rm Gal}) \simeq 1^{\circ} 
\left( \frac{\epsilon_p}{10^{20}{\rm eV}} \right)^{-1} 
\left( \frac{D_{\rm Gal}}{40 {\rm kpc}} \right)^{1/2} 
\left( \frac{l_c}{50 {\rm pc}} \right)^{1/2} 
\left( \frac{B_{\rm turb}}{3 \mu{\rm G}} \right), 
\end{equation}
where $\epsilon_p$ and $D_{\rm Gal}$ are the energy of protons and the typical size of the Galaxy, respectively. This estimation is expected to be larger than the deflection angles of protons by the turbulent field in our simulation, since the turbulent field becomes weaker outside the Galactic disk. Thus, the turbulent component of GMF subdominantly contributes to the total deflection of HEPs \citep*{tinyakov05} and the global deflection pattern of HEPs produced by the coherent components of GMF is not largely disturbed \citep*{yoshiguchi04}.

The spiral component of GMF dominantly affects the deflections of HEPs. This is because the solar system is located in the Galactic disk, where the spiral field is stronger than in the Galactic halo. The spiral field is stronger ($\sim 1.5 \mu$G) than the dipole field ($\sim 0.3 \mu$G) in the vicinity of the solar system. Several examples of the trajectories of HEPs with $10^{19.8}$ eV, which are useful to understand the features of the trajectories in the spiral field, are shown in Fig. \ref{fig:trajectories}. Only the spiral component is taken into account as GMF. The upper panel shows the trajectories of HEPs in the 4 different spiral field models in $x$-$z$ plane. Note that $x$-axis is a line penetrating from the Galactic center to the solar system, and $z$-axis is a line penetrating from the Galactic center to the north Galactic pole. The solar system is located at $x = 8.5$ kpc. The arrival directions of these HEPs at the Earth are the same, $(\ell, b) \sim (180.0^{\circ}, 0.3^{\circ})$. Practically, we calculate the trajectories of oppositely-charged protons injected from the Earth to $(\ell, b) \sim (180.0^{\circ}, 0.3^{\circ})$ and regard these trajectories as those of arriving HEPs. The trajectory in the BS-S model and that in the BS-A model are duplicated because these two models has the same spiral field above the Galactic plane. The trajectory in the AS-S model and that in the AS-A model are in the same situation. The trajectories in the BS models are separated from those in the AS models outside $\sim 15$ kpc because of the existence of the field reversal of the spiral field in the BS models. Once a proton passes over the radius of a field reversal, the deflection direction of the proton is changed into the opposite direction to the direction before. This effect does not occur in the AS models, since the AS models do not have the field reversals of GMF. This effect also allows the total deflection angles of HEPs to be smaller in the BS models than in the AS models. After the oppositely-charged protons propagate away from the Galactic plane, their deflections are smaller because the spiral component of GMF becomes smaller exponentially. Thus, the spiral GMF in the vicinity of the solar system mainly affects the total deflections of HEPs.

The effect of the parity of the spiral component of GMF can be observed in the other panels. The lower left panel is the same as the upper panel, but the arrival directions of HEPs are $(\ell, b) = (180.0^{\circ}, -0.3^{\circ})$. Due to the difference of the parity, the spiral field of the BS-A (AS-A) model has the opposite direction to that of the BS-S (AS-S) model below the Galactic plane. In this panel, the trajectories of oppositely-charged protons are deflected toward the upside of the Galactic plane near the solar system in the cases of GMF models with the S-type parity (BS-S and AS-S), while the trajectories are deflected toward the downside of the Galactic plane for the GMF models with the A-type parity (BS-A and AS-A). Compared to the upper panel, the BS-A and AS-A models show the symmetric deflection pattern.  The lower right panel is another example to show the effect of the parity. HEP trajectories with the same arrival direction, $(\ell, b) = (0.0^{\circ}, -44.8^{\circ})$, are separated by the effects of the parity and field reversals of GMF.

\begin{figure*}
\epsscale{0.85}
\plotone{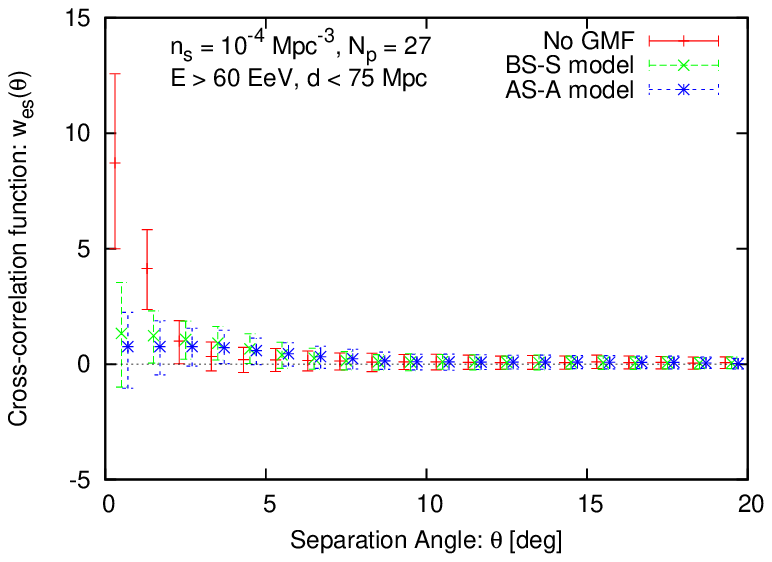}
\plotone{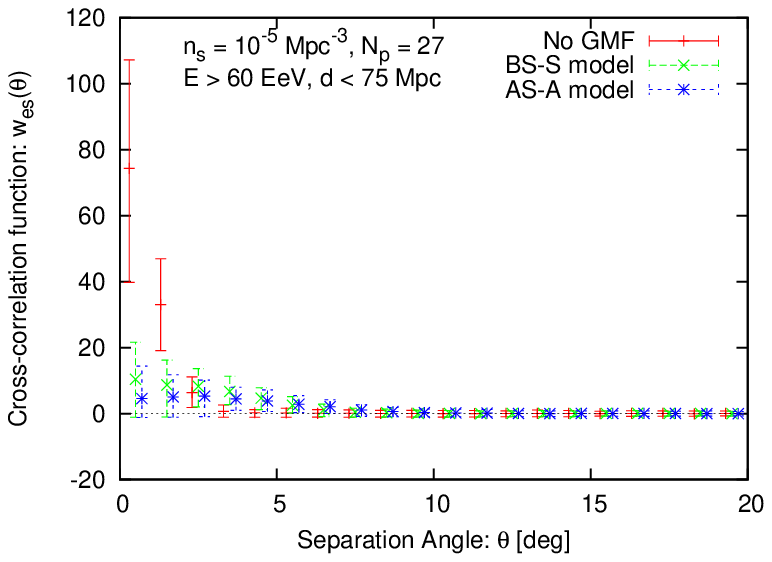}
\caption{Cross-correlation functions between the arrival directions of 27 simulated HEPs with the energy above $6 \times 10^{19}$ eV and the positions of HEP sources with $n_s \sim 10^{-4}$ ({\it left}) and $10^{-5}$Mpc$^{-3}$ ({\it right}) within 75 Mpc. The aperture of the PAO is taken into account. Only the spiral component is considered as GMF; no GMF ({\it green}), the BS-S model ({\it red}), and the AS-A model ({\it blue}).}
\label{fig:gmf0}
\end{figure*}

The deflection directions of HEPs depend on the positions of their sources. Fig. \ref{fig:ad} demonstrates the arrival directions of protons emitted from 8 extragalactic sources at 50 Mpc from the Earth. The positions of the sources are shown by black circles. Only the spiral component of GMF is taken into account. The positions of the sources are artificially set and protons with the energy down to $10^{19}$ eV are considered to see the directions and strengths of the deflections of HEPs. The energy of protons are represented by color in the figure. In order to make these panels, we calculate the trajectories of oppositely-charged protons injected from the Earth, select trajectories which pass the sources, and plot the injection directions of the particles.

The arrival directions of HEPs are basically shifted from the positions of their sources to the directions of low or high latitude \citep*{stanev97,alvarez02,takami08b} and are arranged reflecting their energies because the spiral field is directed parallel to the Galactic plane \citep*{yoshiguchi03}. The deflection directions depend on the structure of the spiral field. Protons emitted from a source with $(\ell, b) \sim (120^{\circ}, 60^{\circ})$ are deflected to lower latitude. We can see that the deflection angles of these protons are smaller in the BS-S model and BS-A model than in the AS-S model and AS-A model, as discussed above. Since the AS-A model and BS-A model have symmetry above and below the Galactic plane, the arrival directions of HEPs from the source and those from a source with $(\ell, b) \sim (120^{\circ}, -60^{\circ})$ are symmetric. This symmetric feature can be seen for HEPs emitted from the other sources. Comparing protons from a source with $(\ell, b) \sim (120^{\circ}, 60^{\circ})$ with those from a source with $(\ell, b) \sim (-60^{\circ}, 60^{\circ})$, for example, we can see the dependence of the deflection angles of protons on source position. The latter protons are less deflected even for $\sim 10^{19.2}$ eV than the former protons. We can also see that protons with the energy of $\sim 10^{19}$ eV are trapped near the Galactic center and are largely scattered.

The arrangement of the deflection directions of protons can be a hint to understand the structure of GMF in the direction of sources \citep*{yoshiguchi03,golup09}. In fact, protons with the energy lower than $\sim 4 \times 10^{19}$ eV are difficult to be a probe of GMF structure, since cosmologically distant sources can significantly contribute to the total flux of HECRs. The background makes the identification of the sources of such low energy protons difficult. On the other hand, HEPs above $\sim 6 \times 10^{19}$ eV is a good tool for this purpose because they can reach the Earth only from nearby sources due to Greisen-Zatsepin-Kuz'min (GZK) effect \citep{greisen66,zatsepin66}. 

\begin{figure*}
\epsscale{0.85}
\plotone{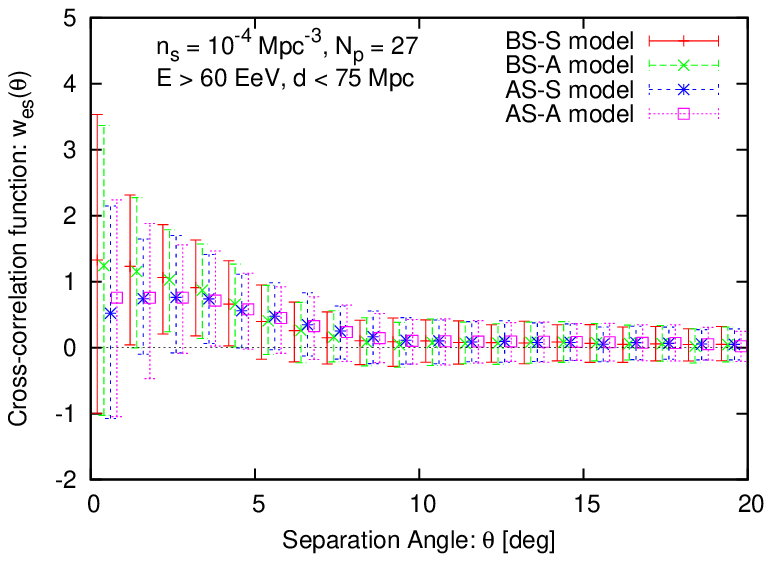}
\plotone{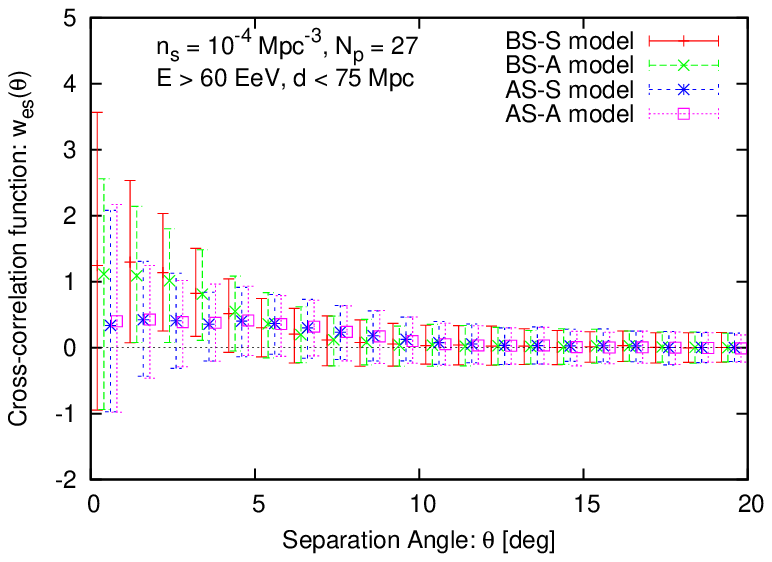}
\caption{Cross-correlation functions of 27 simulated HEPs with the energy above $6 \times 10^{19}$ eV and their sources with $n_s = 10^{-4}$ Mpc$^{-3}$ within 75 Mpc. The BS-S model ({\it red}), the BS-A model ({\it green}), the AS-S model ({\it blue}), and the AS-A model ({\it magenta}) are considered as a spiral component of GMF. The apertures of the PAO ({\it left}) and TA ({\it right}) are taken into account.}
\label{fig:current}
\end{figure*}

Finally, we show how the deflection angles of protons are distributed in the sky as a function of their arrival directions. Fig. \ref{fig:def} shows the distribution of deflection angles of protons with the energy of $10^{19.8}$ eV in equatorial coordinates. Only the spiral component of GMF is taken into account. In all the models, the trajectories of HEPs are highly deflected near the Galactic center. We can also see that the BS models predict the smaller deflection angles of HEPs than the AS models, as discussed in Figs. \ref{fig:trajectories} and \ref{fig:ad}. The difference between the AS models and BS models is noticeable especially in the northern sky, where the Galactic center is not contained. A similar plot taking the dipole field into account can be seen in \cite{takami08b}. The effect of the dipole field appears essentially only in the direction of the Galactic center.

\subsection{Cross-correlation between HEPs and their sources} \label{ccor_sources}

First of all, we discuss the effect of the spiral component of GMF on the cross-correlation function between the arrival directions of HEPs and the positions of their sources. The effects of the dipole and turbulent components of GMF will be considered afterwards. Fig. \ref{fig:gmf0} shows the cross-correlation functions between the arrival directions of simulated HEPs with the energy above $6 \times 10^{19}$ eV and the positions of their sources with $n_s \sim 10^{-4}$ ({\it left}) and $10^{-5}$Mpc$^{-3}$ ({\it right}) within 75 Mpc in simulation. The number of HEPs is set to be 27 events, which is the same as that of the published PAO events. The aperture of the PAO is taken into account. These cross-correlation functions are calculated under 3 different GMF models, i.e., no GMF ({\it red}), the BS-S model ({\it green}), and the AS-A model ({\it blue}). The points represent the averages of the cross-correlation functions over 500 source distributions and the error bars correspond to $1\sigma$ errors, which means that 68\% of the values of the cross-correlation functions over the 500 realizations is contained in the range of the error bars.

\begin{figure*}
\epsscale{0.9}
\plotone{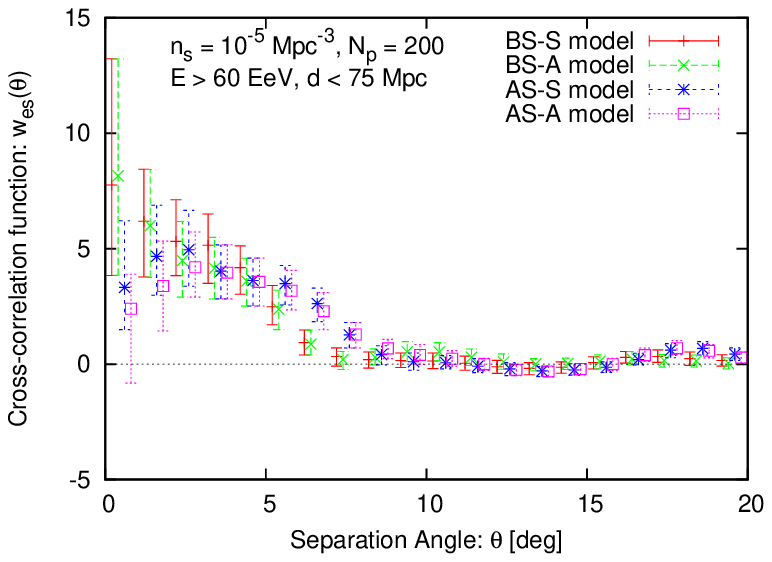}
\plotone{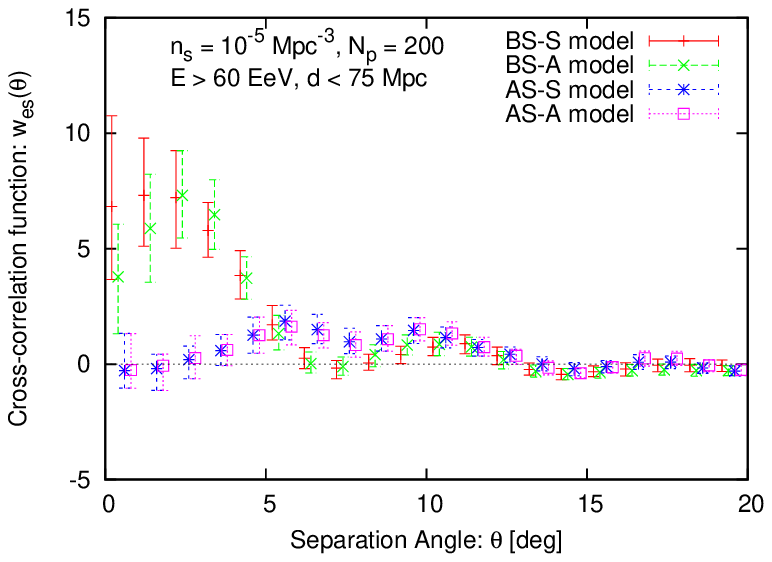}
\caption{Cross-correlation functions between 200 protons simulated from a given source distribution with $n_s \sim 10^{-5}$ Mpc$^{-3}$ and their sources within 75 Mpc. The energy of the protons is above $6 \times 10^{19}$ eV. The apertures of the PAO ({\it left}) and TA ({\it right}) are taken into account. Only the spiral component of GMF is taken into account.}
\label{fig:ccor03}
\end{figure*}

The averages and errors are estimated as follows: we simulate one arrival distribution of a required number of HEPs (27 in this case) for a given source distribution, and then calculate the cross-correlation function between simulated events and sources within 75 Mpc in this simulation. We repeatedly calculate the cross-correlation functions in the same way for 500 different source distributions with the same $n_s$. Then, we calculate the averages of these 500 cross-correlation functions, and the values of the cross-correlation functions at 16\% from the minimum and at 84\% from the maximum of these functions. The latter 2 values are the lower limit and the upper limit of the error bars, respectively. Therefore, the error bars include not only errors due to the finite number of simulated events but also errors originating from the uncertainty of HEP source positions, i.e., the galaxy sampling in our HECR source model (see \S \ref{sources}). Note that the latter error cannot be reduced unless additional conditions are required in the source model. All the cross-correlation functions except for Fig. \ref{fig:ccor03} are plotted in the same way as above.

The cross-correlation functions have a sharp peak at the smallest angular bin for no GMF in the both panels of Fig. \ref{fig:gmf0} because the propagation trajectories of HEPs are straight. The finite accuracy of HECR observatories to determine the arrival directions of HEPs ($\sim 1^{\circ}$) allows the positive values of the cross-correlation functions at the second and third bins. The value of the cross-correlation function at the peak is larger for $n_s = 10^{-5}$ Mpc$^{-3}$ than that for $n_s = 10^{-4}$ Mpc$^{-3}$ because a smaller number of events is enough to trace the positions of sources by smaller number density.  When the BS-S model or the AS-A model of GMF is considered, the positive signals of the cross-correlation functions for no GMF are suppressed. The cross-correlation functions for both models are consistent with no correlation $( w_{\rm es}(\theta) = 0)$ within $1\sigma$ error in both panels except at several angular bins for the BS-S model.

The left panel of Fig. \ref{fig:current} is the same as the left panel of Fig. \ref{fig:gmf0}, but the cross-correlation functions for the BS-A model and the AS-S model are added and the result for no GMF is removed. The cross-correlation functions for the AS models are consistent with no correlation, while those for the BS models have a signal of positive correlation at $3^{\circ}$-$4^{\circ}$, but significance is low. This results from the fact that the BS models predict smaller deflection of HEPs than the AS models. It is suggestive that the angular separation scale of the PAO correlation is $3.1^{\circ}$ \citep*{abraham07,abraham08}. However, note that the analysis of the PAO was based on {\it cumulative} cross-correlation, which is different from {\it differential} cross-correlation used in this study.

\begin{figure*}
\epsscale{0.9}
\plotone{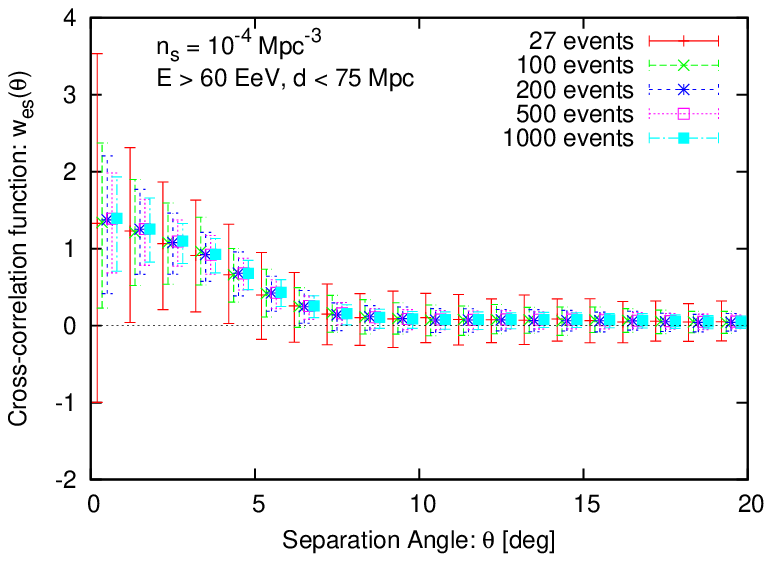}
\plotone{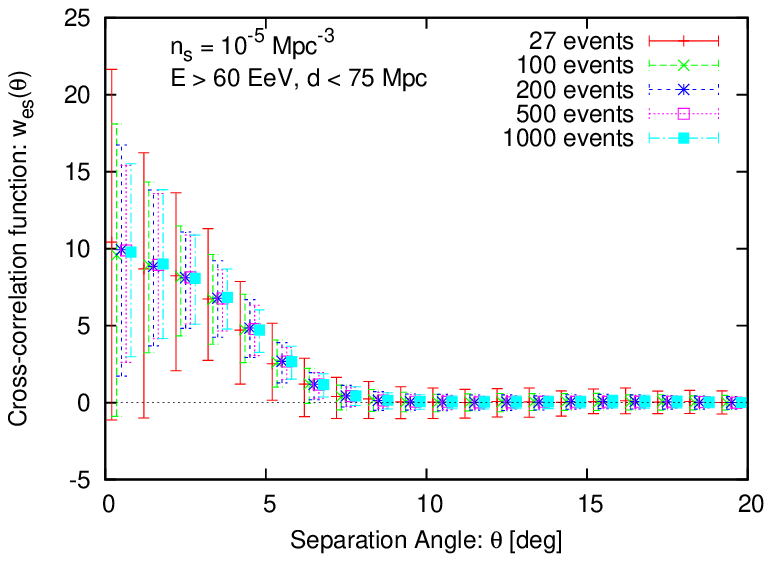}
\caption{Cross-correlation functions between HEPs simulated in the BS-S  model with the energy above $6 \times 10^{19}$ eV and their sources with $n_s = 10^{-4}$ Mpc$^{-3}$ ({\it left}) and $n_s = 10^{-5}$ Mpc$^{-3}$ ({\it right}) within 75 Mpc. The numbers of simulated protons are 27 ({\it red}), 100 ({\it green}), 200 ({\it blue}), 500 ({\it magenta}), and 1000 ({\it light blue}), respectively. The aperture of the PAO is taken into account. 
The error bars start to decrease at $\sim 200$ events.}
\label{fig:convergence}
\end{figure*}

\begin{figure*}
\epsscale{0.9}
\plotone{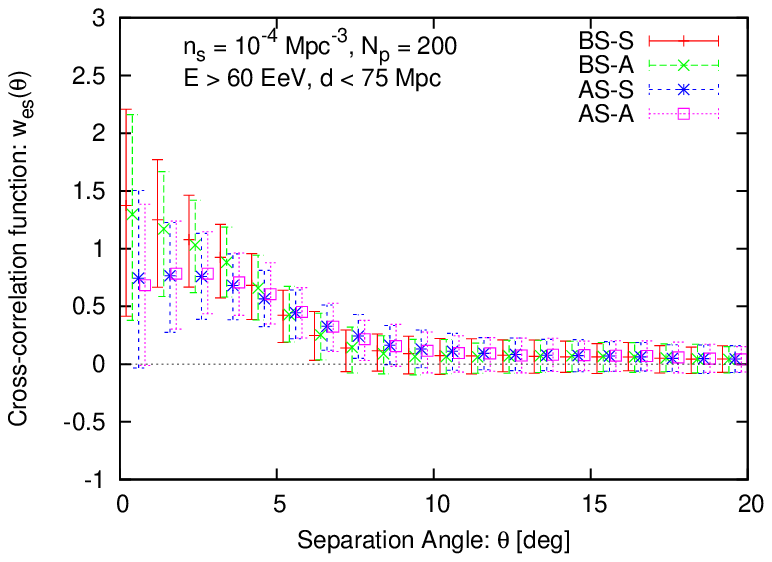}
\plotone{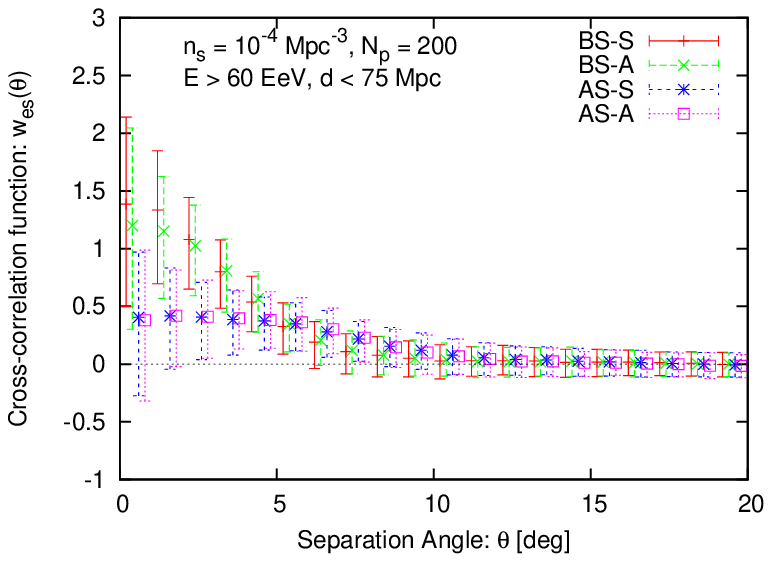}
\caption{Same as Fig. \ref{fig:current}, but the number of simulated HEPs is 200.}
\label{fig:ccor01}
\end{figure*}

\begin{figure*}
\epsscale{0.9}
\plotone{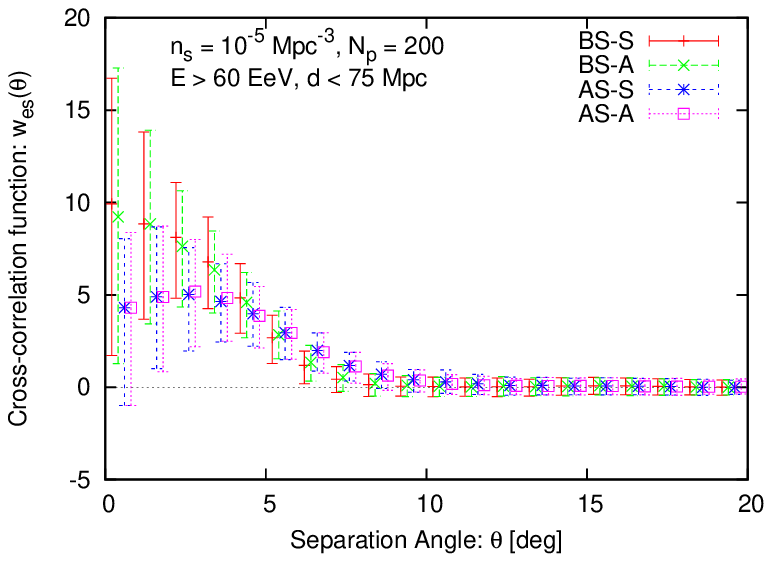}
\plotone{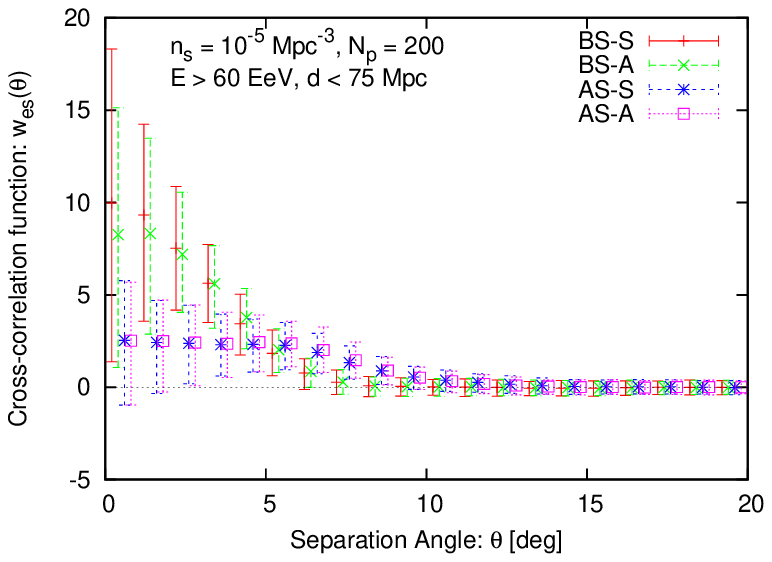}
\caption{Same as Fig. \ref{fig:ccor01}, but for $n_s = 10^{-5}$ Mpc$^{-3}$.} 
\label{fig:ccor02}
\end{figure*}

The right panel of Fig. \ref{fig:current} is the same as the left panel, but the aperture of the TA is taken into account to observe the difference between a cross-correlation signal in the southern sky and that in the northern sky. The number of simulated events is the same as for the PAO, 27. Note that it spends more time than the PAO that the TA accumulate 27 events above $\sim 6 \times 10^{19}$ eV because the aperture of the TA is smaller than that of the PAO. In the northern sky, the AS models predict the cross-correlation functions consistent with no correlation within $1\sigma$ error in all the angular scale. Since the lower limit of the error bars is lower than that for the left panel, detecting the correlation in the northern sky is more difficult than in the southern sky if axisymmetric GMF is realized. Comparing both panels of Fig. \ref{fig:current}, we notice that the cross-correlation functions predicted by the BS-S (AS-S) model is quite similar to those predicted by the BS-A (AS-A). The parity of the spiral component of GMF does not largely affect this statistical quantity, although the parity produce the difference of the deflection directions of HEPs, as seen in Fig. \ref{fig:ad}. The difference between the AS models and the BS models is more clear in the northern sky, as we will show afterwards.

The error bars of the cross-correlation function originate from not only the finite number of events but also the uncertainty of positions of HECR sources (i.e., the galaxy sampling in our source model). The former errors can be decreased by increasing the number of detected events, while the latter errors cannot be reduced unless additional constraints are given in our model. For an ideal case where the number of detected events is infinite, we can consider only the latter errors. In the practical point of view, it is meaningful to estimate the number of events required for saturation of the errors because we can discuss the correlation of HEPs and their sources based on almost only the uncertainty originating from the galaxy sampling after accumulating such number of events. Fig. \ref{fig:convergence} shows the cross-correlation functions between HEPs with the energy above $6 \times 10^{19}$ eV and their sources with $n_s = 10^{-4}$ Mpc$^{-3}$ ({\it left}) and $n_s = 10^{-5}$ Mpc$^{-3}$ ({\it right}) within 75 Mpc. The HEP events are simulated by assuming the BS-S model and the PAO aperture. The numbers of protons considered are 27 ({\it red}), 100 ({\it green}), 200 ({\it blue}), 500 ({\it magenta}), and 1000 ({\it light blue}). When the number of events is small, errors due to the finite number of events are dominated in the total errors. As increasing the number of events, these errors decrease and errors due to the sampling of galaxies become dominant. Since the lengths of error bars for 500 events are not largely different from those for 1000 events, the error bars are well saturated at $\sim 500$ events for both $n_s$. This fact is unchanged for the other GMF models and in the northern sky. The PAO can accumulate $\sim 200$ events above $\sim 6 \times 10^{19}$ eV for 10 years ( 27 events per 1.5 years exposure \citep*{abraham08}). The error bars for 200 events are also sufficiently small. Thus, we consider 200 protons for discussions below.

Fig. \ref{fig:ccor01} shows the cross-correlation functions between 200 simulated HEPs with the energy above $6 \times 10^{19}$ eV and their sources with $n_s = 10^{-4}$ Mpc$^{-3}$ within 75 Mpc for the BS-S model ({\it red}), BS-A model ({\it green}), AS-S model ({\it blue}), and AS-A model ({\it magenta}). The apertures of the PAO ({\it left}) and TA ({\it right}) are taken into account. Compared with Fig. \ref{fig:current}, the error bars become smaller, and the clear signals of positive correlation appear in some angular scale. In addition, we can see that the cross-correlation functions for the BS-S (AS-S) model behave similarly to those for the BS-A (AS-A) model. In the left panel, the cross-correlation functions for the BS models are not consistent with zero (no correlation) up to $7^{\circ}$ at $1\sigma$ error, and therefore the positive correlation appears in this angular scale. On the other hand, the signal of positive correlation for the AS models is in the angular scale from $2^{\circ}$ to $8^{\circ}$. The lower limit of the error bars at small angular scale is lower than that for the BS models. This reflects the fact that the AS models predict the larger deflection of HEPs than the BS models. The right panel shows qualitatively the same result as the left panel. The BS models predict the positive correlation up to $6^{\circ}$, while the cross-correlation functions for the AS models are inconsistent with zero at $3^{\circ}$-$8^{\circ}$. The difference between the cross-correlation functions for the BS models and those for the AS models is larger in the northern sky than in the southern sky. Fig. \ref{fig:ccor02} shows the same figure as for Fig. \ref{fig:ccor01}, but $n_s = 10^{-5}$ Mpc$^{-3}$. The qualitative features of the cross-correlation functions are the same as those in Fig. \ref{fig:ccor01}. The values of the cross-correlation functions are larger than those for $n_s = 10^{-4}$ Mpc$^{-3}$ because of the smaller $n_s$, but the error bars also become larger.

\begin{figure*}
\epsscale{0.85}
\plotone{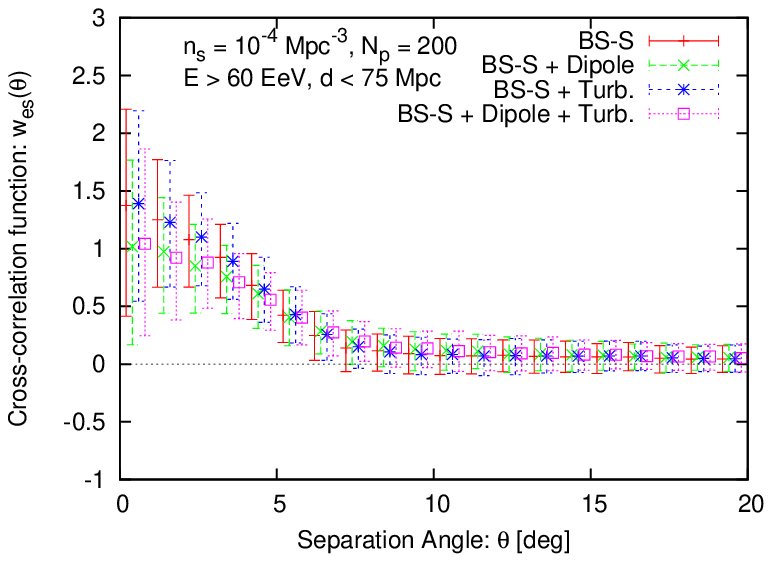}
\plotone{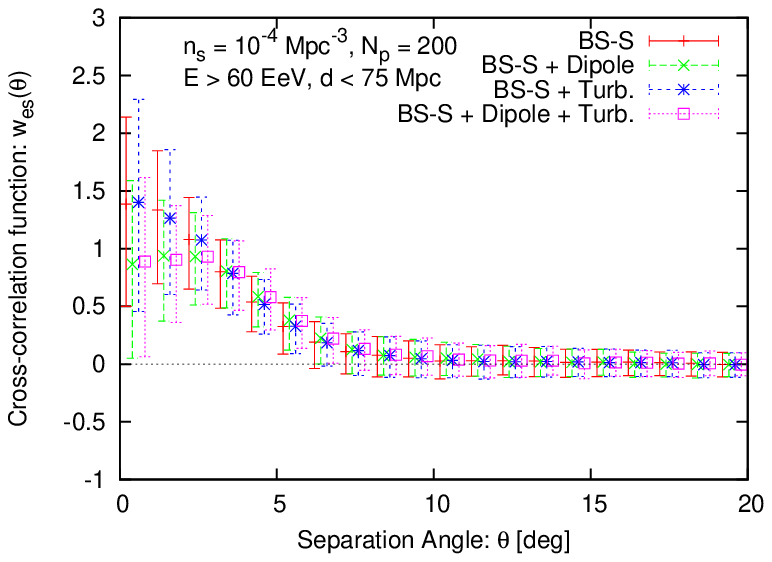}
\plotone{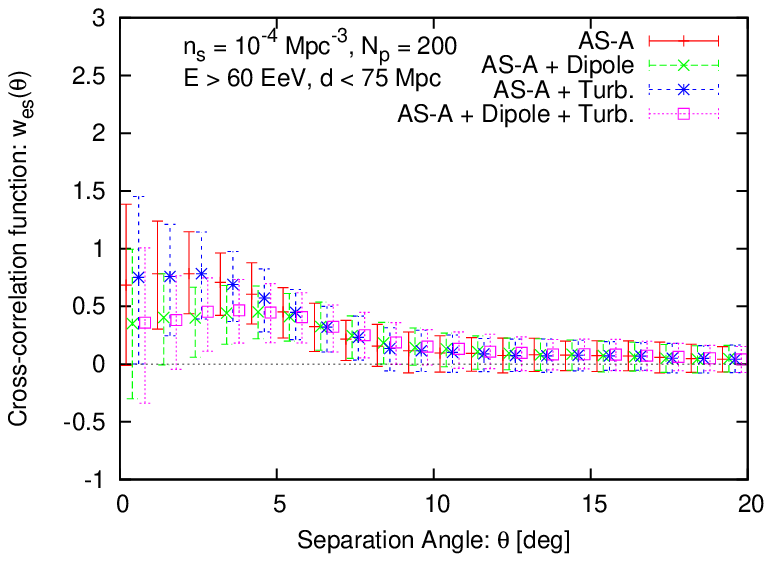}
\plotone{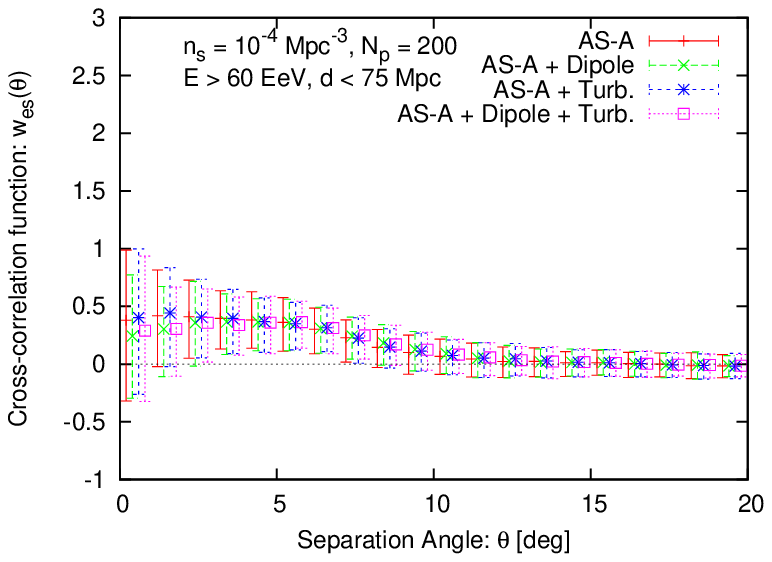}
\caption{Cross-correlation functions between the arrival directions of 200 simulated HEPs with the energy above $6 \times 10^{19}$ eV and the positions of their sources with $n_s = 10^{-4}$ Mpc$^{-3}$ within 75 Mpc. The apertures of the PAO ({\it left}) and the TA ({\it right}) are taken into account. The BS-S model ({\it upper}) and the AS-A model ({\it lower}) are considered as a spiral component of GMF. In addition, the dipole component and the turbulent component of GMF are also considered for comparison.}
\label{fig:ccor_comp}
\end{figure*}

The difference between the BS models and the AS models can be clearly seen in the cross-correlation functions calculated from a given source distribution. In this case, the error bars of the cross-correlation functions are determined by only errors due to the finite number of events, i.e., the error bars can be minimized by accumulating events. Fig. \ref{fig:ccor03} shows the cross-correlation functions based on one specific source distribution with $n_s = 10^{-5}$ Mpc$^{-3}$. The apertures of the PAO ({\it left}) and TA ({\it right}) are considered. The arrival distribution of 200 HEPs with the energy above $6 \times 10^{19}$ eV is realized 500 times. and then the averages and the 68\% error bars of the 500 cross-correlation functions are calculated. For this example, the difference between the cross-correlation functions for the BS models and those for the AS models is clear in the northern sky. The BS models predict the strongly positive correlation within $5^{\circ}$, while the cross-correlation functions are consistent with zero at small angular scale for the AS models. On the other hand, the positive correlation is predicted for all the spiral field models in the southern sky. The angular scale at which the positive correlation appears is larger for the AS models than for the BS models.

Next, we investigate the effect of the dipole component and the turbulent component of GMF on the cross-correlation function. Fig. \ref{fig:ccor_comp} shows the cross-correlation functions between the arrival directions of 200 HEPs with the energy above $6 \times 10^{19}$ eV and the positions of their sources with $n_s = 10^{-4}$ Mpc$^{-3}$ within 75 Mpc. The apertures of the PAO and the TA are taken into account in the left panel and the right panel, respectively. We consider the BS-S model in the upper panel and the AS-A model in the lower panel as a spiral component of GMF ({\it red}). In addition to the spiral component, the dipole component of GMF ({\it green}), the turbulent component of GMF ({\it blue}), and both the dipole and turbulent components ({\it magenta}) are considered for comparison. In all the panel, the dipole field reduces the averages and the positions of the error bars of the cross-correlation functions, and therefore the positive signals of the correlation decreases at small angular scale. On the other hand, the turbulent component of GMF hardly affects the cross-correlation functions.

Finally, we check the dependence of the cross-correlation function on the maximum distance of HEP sources to calculate the cross-correlation functions. So far, the maximum distance has been fixed at 75 Mpc, which is the same distance as the analysis of the PAO \citep*{abraham07,abraham08} ( $z = 0.018$ in the concordance cosmology ). However, this value does not have inevitability, although it is consistent with the GZK scenario, since this value was derived to optimize the correlation signal. Thus, we check the cross-correlation functions for $d < 100$ Mpc and $d < 50$ Mpc.

\begin{figure*}
\epsscale{0.85}
\plotone{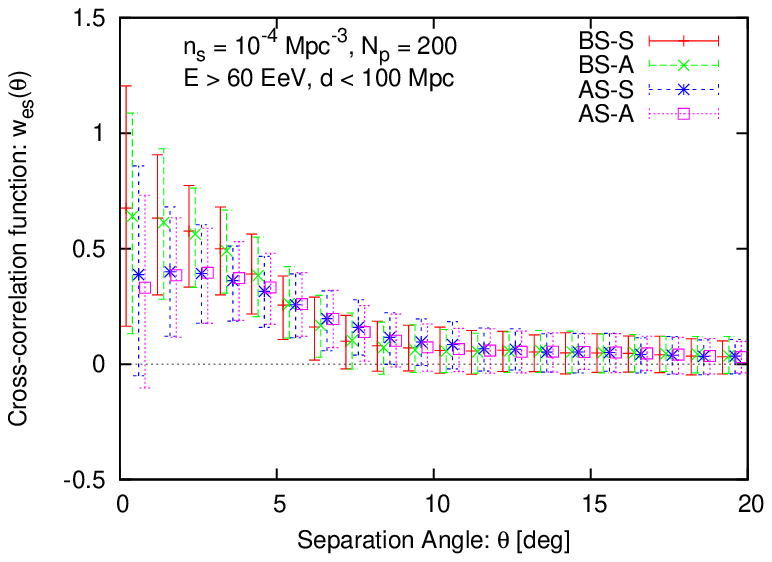}
\plotone{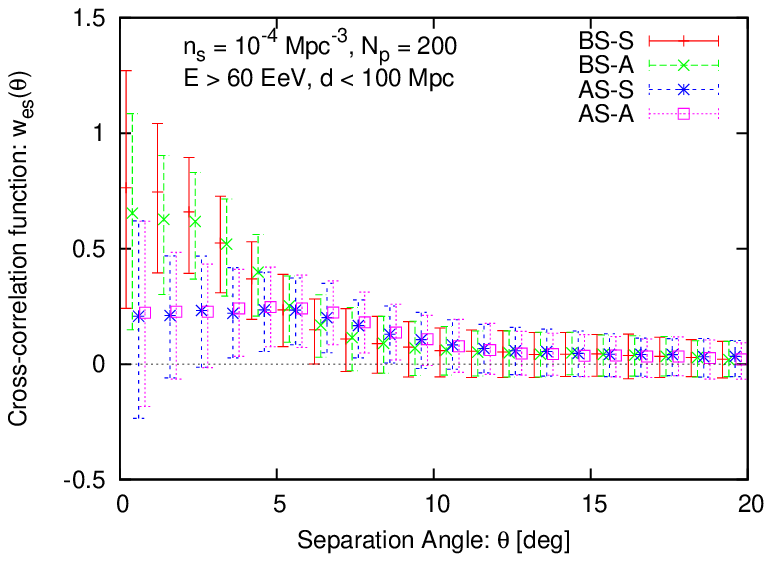}
\plotone{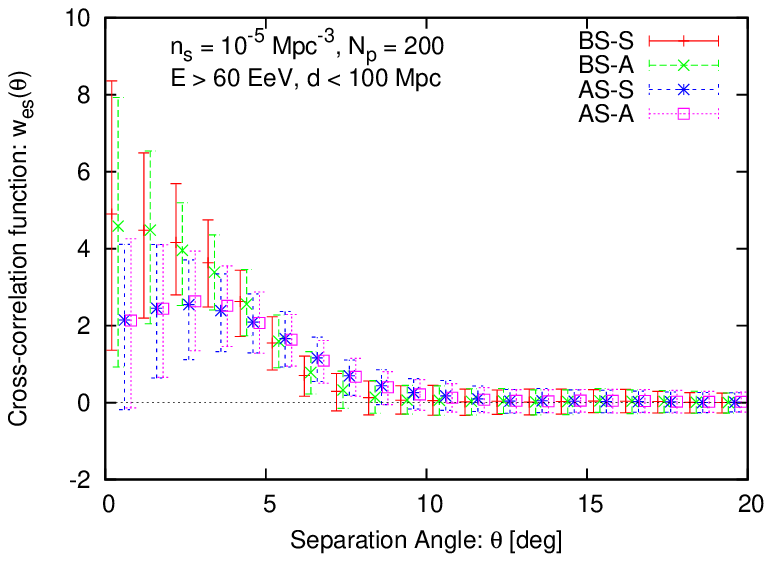}
\plotone{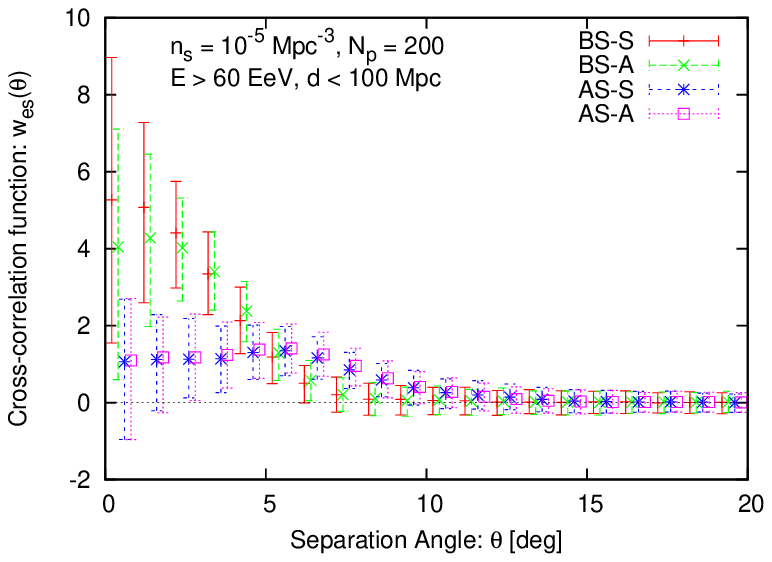}
\caption{Cross-correlation functions of 200 simulated HEPs with the energy above $6 \times 10^{19}$ eV and the positions of their sources with $n_s \sim 10^{-4}$ Mpc$^{-3}$ ({\it upper}) and $n_s \sim 10^{-5}$ Mpc$^{-3}$ ({\it lower}) within 100 Mpc. The BS-S model ({\it red}), the BS-A model ({\it green}), the AS-S model ({\it blue}), and the AS-A model ({\it magenta}) are considered as a spiral component of GMF. The apertures of the PAO ({\it left}) and TA ({\it right}) are taken into account.}
\label{fig:ccor04}
\end{figure*}

\begin{figure*}
\epsscale{0.85}
\plotone{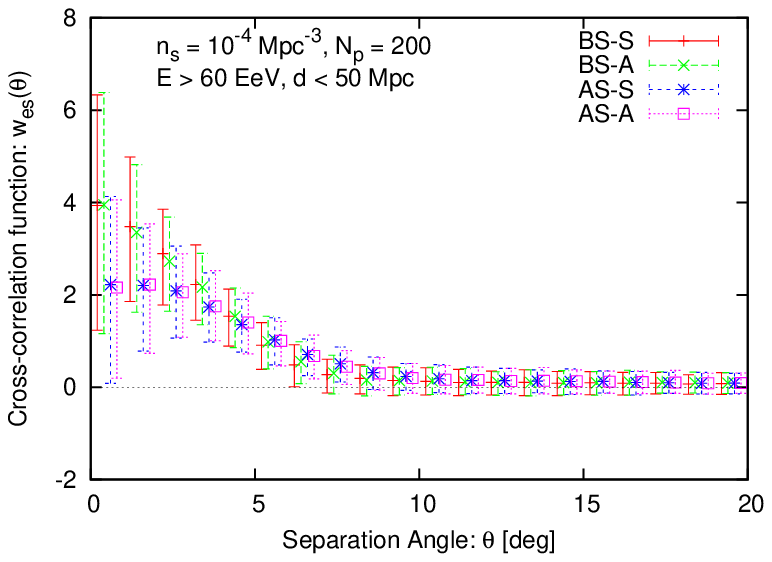}
\plotone{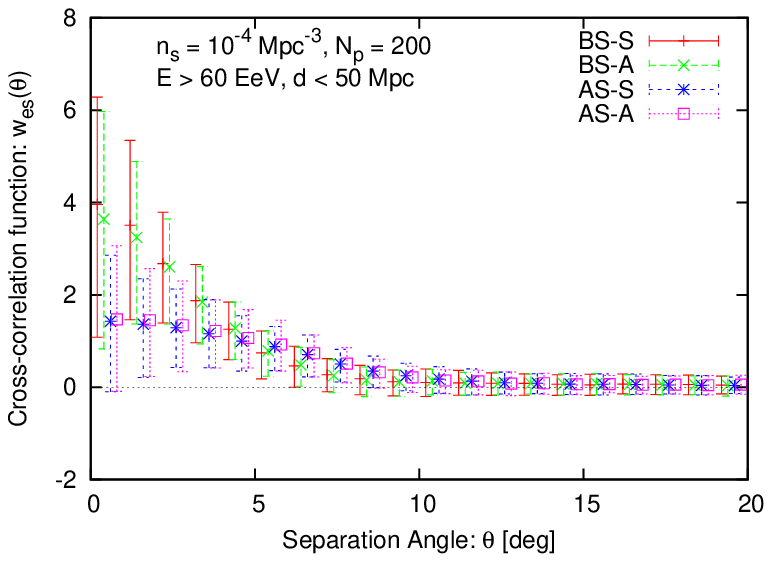}
\plotone{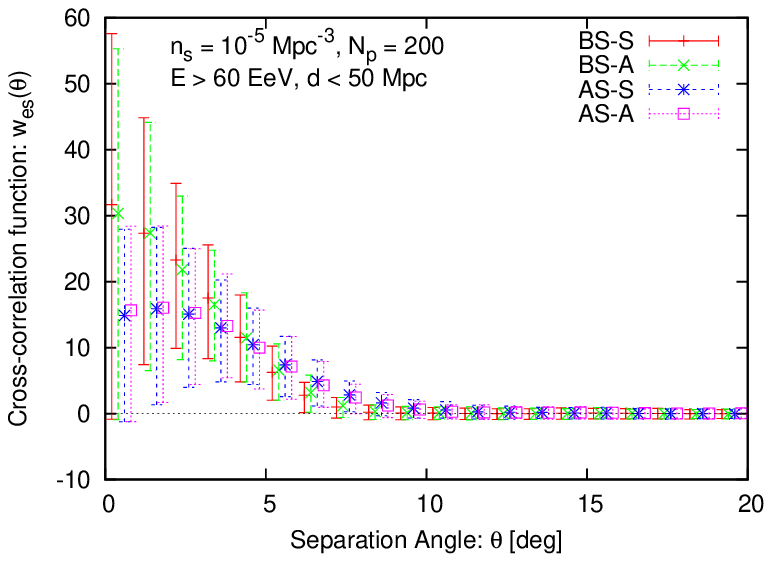}
\plotone{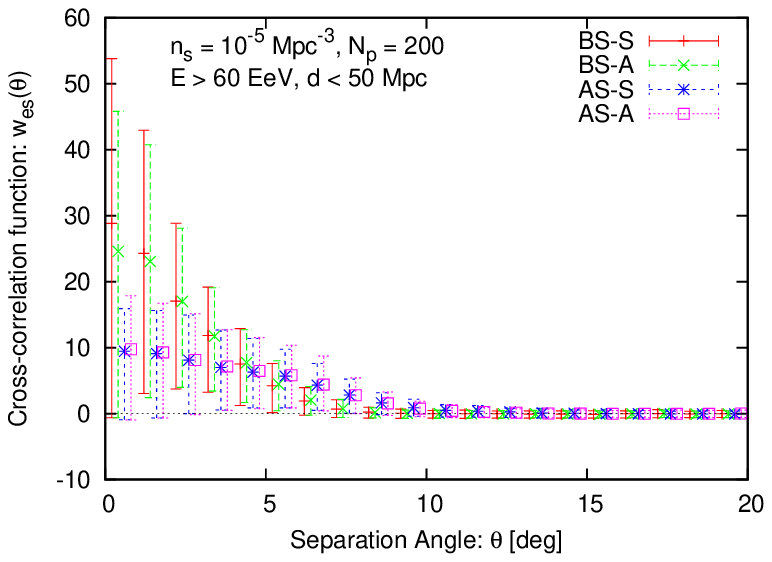}
\caption{Same as Fig. \ref{fig:ccor04}, but for HEP sources within 50 Mpc.}
\label{fig:ccor05}
\end{figure*}

Fig. \ref{fig:ccor04} shows the cross-correlation functions of 200 HEPs with the energy above $6 \times 10^{19}$ eV and their sources with $n_s \sim 10^{-4}$ Mpc$^{-3}$ ({\it upper}) and $n_s \sim 10^{-5}$ Mpc$^{-3}$ ({\it lower}) within 100 Mpc. The apertures of the PAO ({\it left}) and TA ({\it right}) are taken into account. Compared to Figs. \ref{fig:ccor01} and \ref{fig:ccor02}, the qualitative features of the cross-correlation signals are the same. However, the values of the averages and the positions of the error bars become smaller than those in Figs. \ref{fig:ccor01} and \ref{fig:ccor02}. The lengths of the error bars also become smaller. These are because the number of sources considered to calculate the cross-correlation functions increases due to increasing the maximum distance of the sources for calculating the cross-correlation functions. The large number of sources decreases the fluctuation of galaxy sampling. Fig. \ref{fig:ccor05} shows the same as Fig. \ref{fig:ccor04}, but the maximum distance of HEP sources considered to calculate the cross-correlation functions is 50 Mpc. These cross-correlation functions also have the same features qualitatively as those in Fig. \ref{fig:ccor04}. Contrary to Fig. \ref{fig:ccor04}, the values of the cross-correlation functions are larger than those in Figs. \ref{fig:ccor01} and \ref{fig:ccor02} because of the smaller number of sources considered to calculate the cross-correlation functions. The small number of sources also produces the large fluctuations of the cross-correlation functions. Thus, the qualitative features of the positive correlation are not dependent of $d$ if $d < 100$ Mpc. 

\subsection{Angular Distance Sensitive to Positive Correlation} \label{ccor_best}

In \S \ref{ccor_sources}, we investigated the appearance of the positive correlation between HEPs and their sources based on a general source model in which HEP sources are distributed with being comparable to galaxy distribution. The consideration of 500 source distributions allows us to conclude that the positive correlation can appear at angular separation scale less than $10^{\circ}$ even considering the effect of GMF, but this angular scale depends on the modelling of GMF. On the other hand, the positive correlation between HECRs and their source candidates is generally tested by comparing the experimental data of HECR events to random distribution assuming one source candidates distribution \citep*{abraham07,abraham08,kashti08,george08,ghisellini08,takami09c}. In the analysis of the PAO \citep*{abraham07,abraham08}, the energy threshold of HECRs, the maximum distance of source candidates to investigate the correlation, and the angular separation scale between the arrival directions of HECRs and the positions of source candidates were regarded as parameters, and the parameters were constrained so that the significance of excess events around source candidates over random event distribution was maximized. (In fact, the PAO constrained these parameters by the first half of detected events, and then tested the correlation under the same parameter est by using the second half of detected events.) In this section, we estimate the angular scale at which the significance of the positive correlation between HECRs and their sources over random event distribution is maximized. The energy threshold of HECRs and the maximum distance of sources to investigate the correlation are fixed to $6 \times 10^{19}$ eV and 75 Mpc, respectively.

For random event distribution, the number of events which falls into circles with the angular radii $\theta$, $j$, follows binomial distribution, 
\begin{eqnarray}
Bi\left( N,p_{\rm ran}(\theta);j \right) = \left( 
\begin{array}{c}
N \\ j 
\end{array}
\right) \left\{p_{\rm ran}(\theta)\right\}^j \left( 1 - \left\{p_{\rm ran}(\theta)\right\} \right)^{N - j}, 
\end{eqnarray}
where $N$ is the total number of events. $p_{\rm ran}(\theta)$ is equivalent with the fraction of the sky (weighted by the exposure) defined the regions at angular separation less than the angular scale from the selected source candidates \citep*{abraham07,abraham08}. Even when the effect of GMF is taken into account, we confirmed that the number of events within $\theta$ from the source candidates follows binomial distribution. The effect of GMF is included in a parameter $p_{\rm sim}(\theta)$, but $p_{\rm sim}(\theta)$ is no longer the fraction of the sky regions.

The most sensitive angular scale to the positive correlation can be found as follows. A source distribution is considered. First of all, we simulate ($N=$) 200 HEP events and count the number of events within $\theta$ from the positions of sources, which is written as $n_{\rm sim}$. We repeat this step many times (10000 times in this study) and make the histogram of $n_{\rm sim}$. Then, we evaluate $p_{\rm sim}(\theta)$ by fitting this histogram by binomial distribution, $Bi\left( 200,p_{\rm sim}(\theta);j \right)$. We can also estimate $p_{\rm ran}(\theta)$ from random distribution. The number of events inside circles centered at the source positions with radii less than $\theta$ is $n_{\rm ran}$. The probability that the positive correlation is realized corresponds to the probability that ${\tilde n} \equiv n_{\rm sim} - n_{\rm ran}$ is positive. The probability distribution following ${\tilde n}$ is calculated by the convolution of the two binomial distributions, 
\begin{eqnarray}
&& P\left(N, p_{\rm sim}(\theta), p_{\rm ran}(\theta); {\tilde n}\right) \nonumber \\
&=& \left\{ p_{\rm sim}(\theta) \right\}^{\tilde n} 
\left\{ 1 - p_{\rm sim}(\theta) \right\}^{N - \tilde n} 
\left\{ 1 - p_{\rm ran}(\theta) \right\}^N \nonumber \\
&& \sum_m \left( 
\begin{array}{c}
N \\ {\tilde n} + m
\end{array}
\right) \left( 
\begin{array}{c}
N \\ m
\end{array}
\right) 
\left[ \frac{p_{\rm sim}(\theta) p_{\rm ran}(\theta)}
{\left( 1 - p_{\rm sim}(\theta) \right) \left( 1 - p_{\rm ran}(\theta) \right)} 
\right]^m, 
\end{eqnarray}
where $m$ is summed over $0 \leq m \leq N - {\tilde n}$ for $n_{\tilde n} \geq 0$ and $-{\tilde n} \leq m \leq N$ for ${\tilde n} < 0$. Since this is a function of $\theta$ through $p_{\rm sim}(\theta)$ and $p_{\rm ran}(\theta)$, we can estimate $\theta$ at which $\sum_{{\tilde n}=-N}^{{\tilde n}=0} P\left(N, p_{\rm sim}(\theta), p_{\rm ran}(\theta); {\tilde n} \right) $ is minimized, $\theta_{\rm best}$. $\sum_{{\tilde n}=-N}^{{\tilde n}=0} P\left(N, p_{\rm sim}(\theta), p_{\rm ran}(\theta); {\tilde n} \right)$ gives the probability that the number of HEP events correlating their sources does not exceed the number of random events correlating the source positions. The values of $p_{\rm sim}(\theta)$ and $p_{\rm ran}(\theta)$ depend on source distribution because the apertures of experiments are not uniform and the effect of GMF depends on the arrival directions of HEPs. Thus, we calculate $\theta_{\rm best}$ for 500 source distributions for each $n_s$.

\begin{figure*}
\epsscale{0.85}
\plotone{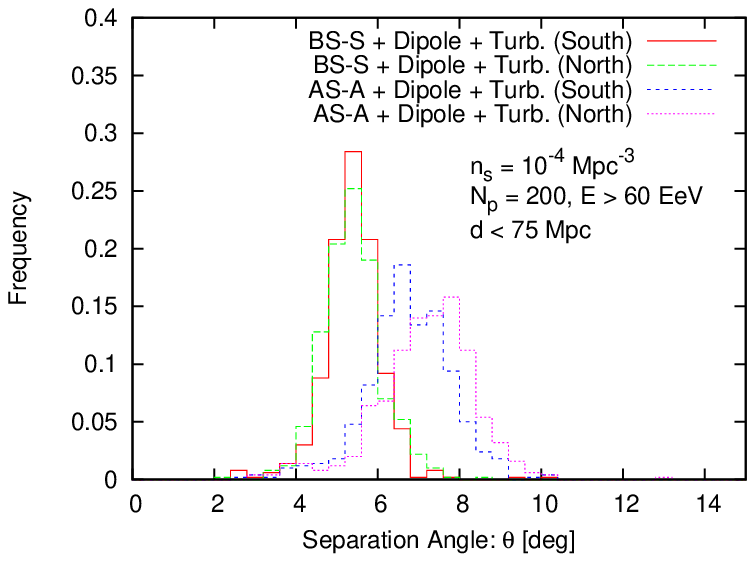}
\plotone{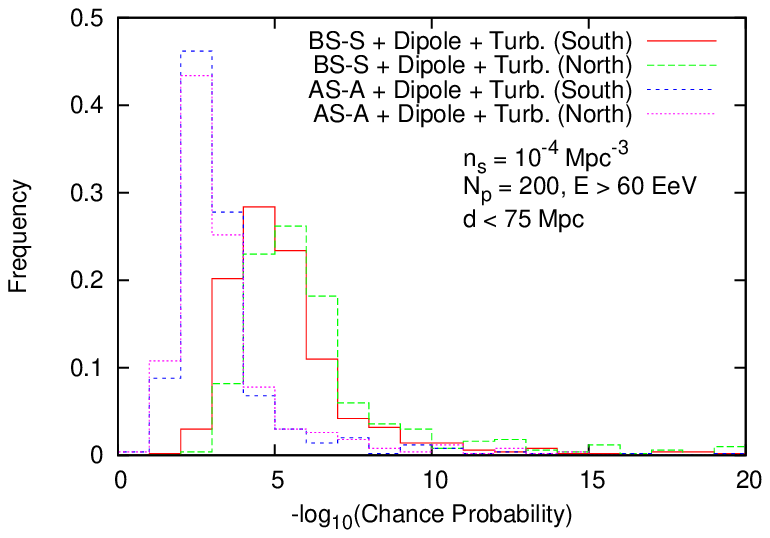}
\plotone{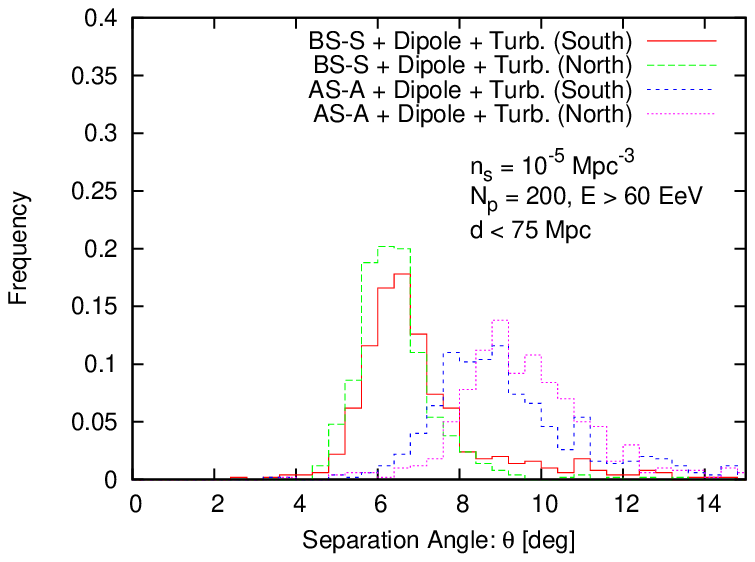}
\plotone{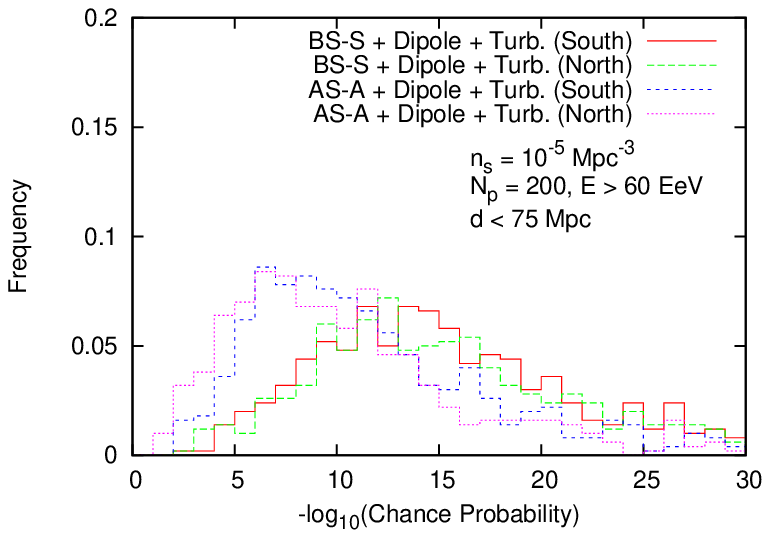}
\caption{({\it left}): Frequency distributions of the separation angle between 200 simulated HEPs with the energy above $6 \times 10^{19}$ eV and their sources with $n_s = 10^{-4}$ Mpc$^{-3}$ ({\it upper}) and $n_s = 10^{-5}$ Mpc$^{-3}$ ({\it lower}) within 75 Mpc which gives maximal significance against random event distribution, $\theta_{\rm best}$, over 500 source distributions. The apertures of the PAO ({\it South}) and the TA ({\it North}) are taken into account. The BS-S model and AS-A model are considered as a spiral component of GMF. The dipole component and turbulent component are also taken into account. ({\it right}): Frequency distribution of the probability that the number of simulated events inside circles with the radii of $\theta_{\rm best}$ centered at the position of their sources does not exceed to that for randomly distributed events. In other words, these are the distributions of the probability that the positive correlation does not occur. All the parameters used are the same as those in the corresponding left panels.}
\label{fig:thbest1}
\end{figure*}

The left panels of Fig. \ref{fig:thbest1} shows the distributions of $\theta_{\rm best}$ for $n_s = 10^{-4}$ Mpc$^{-3}$ ({\it upper}) and $10^{-5}$ Mpc$^{-3}$ ({\it lower}). The BS-S and AS-A models are considered as a spiral component of GMF. The dipole component and turbulent component are also taken into account. The apertures of the PAO ({\it South}) and the TA ({\it North}) are taken into account. The peak of $\theta_{\rm best}$ distributions strongly depend on the spiral shape of GMF. In the case of $n_s = 10^{-4}$ Mpc$^{-3}$, the peak of $\theta_{\rm best}$ distribution is $\sim 5^{\circ}$ for the BS-S model, while that is $6^{\circ}$ - $8^{\circ}$ for the AS-A model. The cases of $n_s = 10^{-5}$ Mpc$^{-3}$ predict larger $\theta_{\rm best}$ at the peak; $\sim 6^{\circ}$ for the BS-S model and $8^{\circ}$ - $10^{\circ}$ for the AS-A model. The peak positions are smaller for $n_s = 10^{-4}$ Mpc$^{-3}$ than that for $n_s = 10^{-5}$ Mpc$^{-3}$. If the number density of sources is large, source distribution projected onto the sky is dense. In this case, the source closest to the arrival direction of a HEP is sometimes not a real source. Thus, the separation angles between HECRs and sources tend to be underestimated for larger $n_s$. The difference between $\theta_{\rm best}$ distribution in the northern sky and that in the southern sky can be seen for the AS-A model. The peak position in the southern sky is $\sim 2^{\circ}$ smaller than that in the northern sky, although the 2 distributions overlap.

The right panels of Fig. \ref{fig:thbest1} shows the distribution of the probability that ${\tilde n} \leq 0$ by chance at $\theta_{\rm best}$, i.e., $\sum_{{\tilde n}=-N}^{{\tilde n}=0} P\left(200, p_{\rm sim}(\theta_{\rm best}), p_{\rm ran}(\theta_{\rm best}); {\tilde n} \right)$ Parameters and GMF models used are the same as those in the corresponding left panels. In both $n_s$, the BS-S model shows smaller chance probability than the AS-A model. Almost all the source distributions predict the chance probability less than $10^{-3}$, which corresponds to the significance of the positive correlation more than $\sim 3\sigma$ in the context of Gaussian distribution, for the BS-S model. On the other hand, the AS-A model can predict larger chance probability because of the larger deflection of HEPs. Less than 5\% of source distributions predicts chance probability larger than $10^{-3}$ for $n_s = 10^{-5}$ Mpc$^{-3}$ . The situation is worse for $n_s = 10^{-4}$ Mpc$^{-3}$ because of the larger number density. About $50\%$ of source distributions predicts chance probability larger than $10^{-3}$. In other words, about $50\%$ of source distributions does not produce the positive correlation at the significance more than $3\sigma$.

\begin{figure*}
\epsscale{0.85}
\plotone{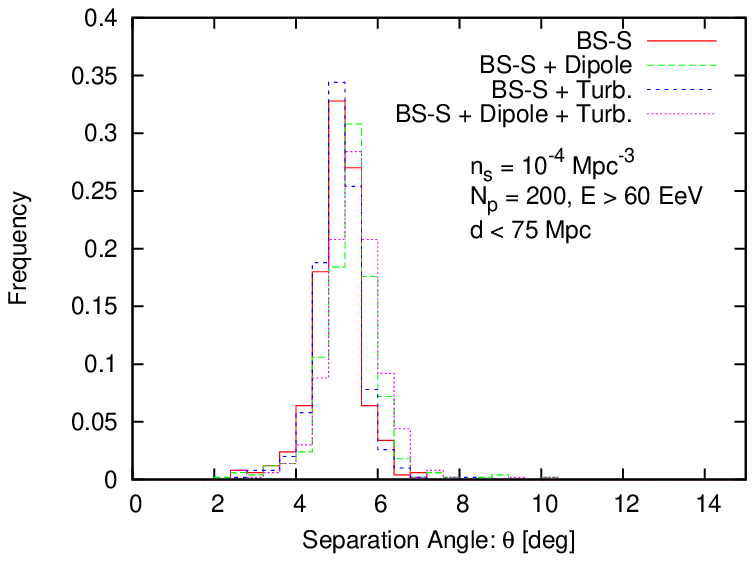}
\plotone{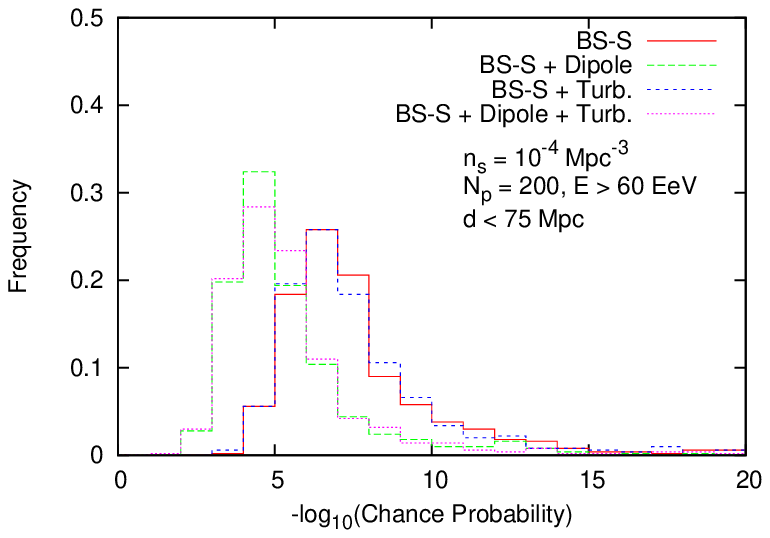}
\plotone{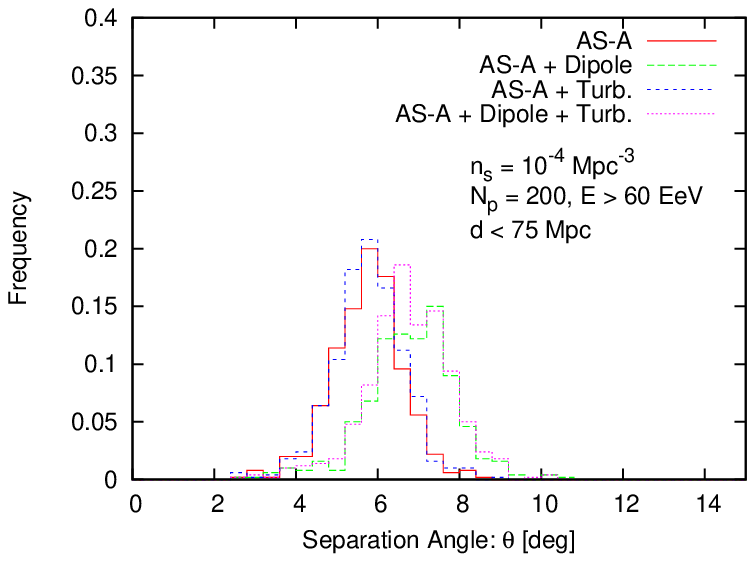}
\plotone{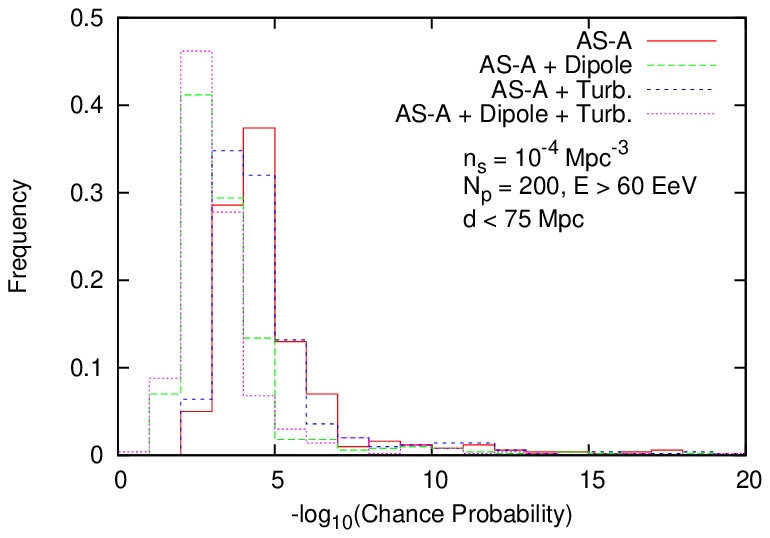}
\caption{({\it left}): Frequency distributions of the separation angle between 200 simulated HEPs with the energy above $6 \times 10^{19}$ eV and their sources with $n_s = 10^{-4}$ Mpc$^{-3}$ within 75 Mpc which gives maximal significance against random event distribution, $\theta_{\rm best}$, over 500 source distributions. The aperture of the PAO is taken into account. The BS-S model ({\it upper}) and AS-A model ({\it lower}) are considered as a spiral component of GMF ({\it red}). In addition to the spiral component, the dipole component of GMF ({\it green}), the turbulent component of GMF ({\it blue}), and both the dipole and turbulent components of GMF ({\it magenta}) are taken into account. ({\it right}): Frequency distribution of the probability that the number of simulated events inside circles with the radii of $\theta_{\rm best}$ centered at the position of their sources does not exceed to that for randomly distributed events. In other words, these are the distribution of the probability that the positive correlation does not occur. All the parameters used are the same as those in the corresponding left panels.}
\label{fig:thbest2}
\end{figure*}

The dependence of these values on the dipole component and turbulent component of GMF is seen in Fig. \ref{fig:thbest2}. $n_s = 10^{-4}$ Mpc$^{-3}$ is considered in this figure. The left panels of Fig. \ref{fig:thbest2} show $\theta_{\rm best}$ distributions for the BS-S model ({\it upper}) and the AS-A model ({\it lower}). In each panel, we consider the 4 configurations of GMF, only a spiral component, spiral component plus dipole component, spiral component plus turbulent component, and all the 3 components. The turbulent component hardly affects $\theta_{\rm best}$ distribution contrary to the dipole component. The dipole component moves the $\theta_{\rm best}$ distribution to the right side by $\sim 0.5^{\circ}$ for the BS-S model and $\sim 1^{\circ}$ for the AS-A model. Thus, additional regular components of GMF to the spiral field can significantly affect the angular scale of positive correlation. This effect of the dipole field is also observed in the right panels of Fig. \ref{fig:thbest2}. The dipole component of GMF increases chance probability. If the dipole field does not exist, only $5\%$ of source distributions produce chance probability larger than $10^{-3}$.

One often uses a penalty factor to estimate real significance when the experimental data of HECRs is analyzed \citep*{tinyakov01,tinyakov04,finley04}. The small number of HECRs detected at present indicates low significance of the correlation even if the correlation is true. Therefore, the significance must be accurately estimated to confirm the correlation. In order to estimate the significance, one scans over parameters (e.g., the maximum distance (redshift) of sources, the energy threshold of HECRs, angular separation, and so on) and finds the parameter set which gives the largest significance of the positive correlation. However, this significance is not real significance because this is optimized to experimental data, which is only one realization of HECR generation. Thus, one estimates a penalty factor originating from the optimization of the parameter sets and corrects the significance. The need of the penalty factor originates from the fact that experimental data is only one realization. However, we can make many realizations of HEP events by simulation in this study and can estimate a real significance, as performed above. Thus, we do not need to consider a penalty factor in this study. 

\section{Discussion \& Conclusion} \label{conclusion}

In this study, we considered the effect of GMF on the expected positive correlation between the arrival distribution of HEPs with the energy above $6 \times 10^{19}$ eV and the positions of their sources by using simulations. We found that the positive correlation can be detected at angular separation scale less than $10^{\circ}$ even considering plausible GMF models when 200 HEPs above $6 \times 10^{19}$ eV are detected.  In addition, the dependence of the positive correlation on GMF models was examined. The trajectories of HEPs are deflected more largely in the AS models than in the BS models because of the lack of field reversals. This fact shows that the signal of the positive correlation is weaker for the AS models than for the BS models. The difference between the AS models and the BS models is prominent especially in the northern sky. The parity of the spiral field does not affect the correlation signals. The dipole field, which is a model of the vertical component of GMF, can subdominantly contribute to the total deflections of HEPs, while the turbulent component of GMF hardly affects the deflections of HEP trajectories. Furthermore, we estimated the angular scale at which the significance of the positive correlation over random event distribution is maximized. The most sensitive angular scale to the positive correlation depends on the spiral shape of GMF; $\theta_{\rm best} \sim 5^{\circ}$ for the BS models and $\sim 7^{\circ}$ for the AS models. The significance of these positive correlation well exceed $\sim 3\sigma$ in almost all the source distributions with $n_s = 10^{-4}$ and $10^{-5}$ Mpc$^{-3}$, but $\sim 50\%$ of source distributions with $n_s = 10^{-4}$ Mpc$^{-3}$ predicts the positive correlation at the significance less than $3\sigma$. This problem can be avoided by neglecting the dipole component. The dipole component contributes to $\theta_{\rm best}$ by $0.5^{\circ}$ - $1^{\circ}$.

We assumed a dipole field as a vertical component of GMF, but the existence of a dipole magnetic field has not been not confirmed. There is large uncertainty on a vertical component of GMF at present. Galactic wind can be also a source of the vertical component of GMF, can predict vertical magnetic field different from the dipole shape \citep*{brandenburg93}. Recent soft X-ray observations have indicated the existence of Galactic wind in the Galaxy \citep{everett08}. Furthermore, high-sensitivity observations of several edge-on galaxies with galactic wind have revealed so-called X-shaped magnetic field in their halo \citep{krause07,beck09}. Although it is uncertain that the Galaxy has such non-dipole vertical component of GMF, such a component could contribute to the deflections of HEPs if exists.

In addition to GMF, IGMF can also contribute to the total deflections of the trajectories of HEPs. However, since our current understanding on IGMF is poor, the deflections highly depend on IGMF modelling. A numerical simulation on cosmological structure formation in local Universe showed a magnetic structure is highly concentrated following matter distribution \citep{dolag05}. Since cross-sectional area occupied by strong IGMF is small, HEPs are deflected less than $1^{\circ}$ except in the directions of clusters of galaxies. A simple modelling of IGMF by \cite{takami06} also predicted small deflection angles of HEPs and showed that HEP sources are unveiled by the arrival directions of HEPs within a few degree \citep{takami08a}. These 2 IGMF models are based on realistic matter distribution of local Universe constructed by the IRAS catalog of galaxies \citep*{saunders00}. On the other hand, hydrodynamical simulations on cosmological structure formation by \cite{sigl04} and \cite{das08} predicted relatively strong IGMF in filamentary structures up to 10nG. The large cross-sectional area occupied by filamentary structures with $\sim 10$ nG fields makes HEPs be deflected significantly. Note that these IGMF models have uncertainty on observer's positions because their resultant matter distributions do not correspond to local Universe actually observed. The IGMF model of \cite{sigl04} deflects the trajectories of HEPs by $\sim 20^{\circ}$ on average even for $10^{20}$ eV. Thus, the large amount of information on the positions of their sources is lost. However, the correlation between HEPs and matter distribution is not always lost even under strong IGMF models. \cite{kotera08} pointed out the possibility of spurious correlation based on another IGMF modelling. In this scenario, HECRs are scattered at strongly magnetized regions far from the positions of their sources and then reach the Earth. In this case, we can observe the correlation between HEPs and last scattering regions, which trace matter distribution. \cite{ryu09} also showed similar possibility under the IGMF modelling of \cite{das08}. As discussed above, the uncertainty of IGMF modellings does not allow us to conclude that the correlation between the arrival directions of HEPs and their sources can be observed. The future projects of all-sky survey of Faraday rotation measurement, like Square Kilometer Array \footnote{http://www.skatelescope.org/} will give us useful information on IGMF in local Universe.

Although we assumed that all the sources emit HEPs with the same power, it is possible that the power of HECR emission depends on sources. In addition, the anisotropic emission of HECRs, e.g., HECR generation in relativistic jets, can change the contribution of a source to the total flux of HECRs because the trajectories of HECRs which can reach the Earth are limited. This anisotropy can be also generated by magnetic field in the cluster of galaxies even if sources emit HECRs isotropically \citep*{dolag09}.

Heavy nuclei in HECRs are crucial for observing the correlation between HECRs and their sources due to their larger electric charge. The propagation trajectories of heavy nuclei are deflected more strongly than those of protons. In this study, we focused on the proton component of HECRs. However, the recent report of the PAO showed that HECRs include a large fraction of the heavier nuclei based on $< X_{\rm max} >$ and the deviations of $X_{\rm max}$ measurements \citep{bellido09}, while the HiRes reported proton-dominated composition \citep{belz09}. On the other hand, the PAO simultaneously reported that the correlation between the arrival directions of HECRs and the positions of nearby astrophysical objects \citep{abraham07,abraham08,hague09}. In order to consider the consistency between the result composition measurement and the correlation, we briefly investigate the propagation of heavy nuclei in GMF.

Heavy nuclei are disintegrated through interactions with CMB photons at the highest energies. Since photo-disintegration produces lighter nuclei than original nuclei, different nuclei are contained in detected HECRs even if only iron nuclei are generated at sources (e.g., \cite{allard05}). Nevertheless, we consider pure iron composition to see the effect of GMF on heavy nuclei clearly. A significant fraction of the PAO events has arrival directions directed to Centaurus A (Cen A), the nearest radio-loud AGN ($\sim 3.4$ Mpc). Since the distance of Cen A is less than the mean free path of photo-disintegration of irons at $\sim 10^{20}$ eV, irons can arrive at the Earth from this source without significant photo-disintegration. Also, irons can be most easily accelerated up to $10^{20}$ eV following the Hillas criterion \citep*{hillas84}. Although pure iron composition is an extreme case, a significant fraction of irons can be expected if Cen A is really one of the HECR sources.

Fig. \ref{fig:traFe} shows examples of the trajectories of irons with the energy of $10^{19.8}$ eV projected onto a plane perpendicular to the Galactic disk and including the Galactic center and the Earth. Their arrival directions are $(\ell, b) = (180.0^{\circ}, 0.3^{\circ})$. This figure corresponds to the lower right panel of Fig. \ref{fig:trajectories}, but the dipole component and the turbulent component of GMF are also taken into account. The trajectories are complicated and do not follow a simple picture explained at Fig. \ref{fig:trajectories}. The red and magenta lines are trapped by strong field including the turbulent component. In any case, the directions of sources are quite different from the arrival directions of irons. The propagation of iron nuclei in GMF is studied in detail by \cite{giacinti10}.

\vspace{0.3cm}
\centerline{{\vbox{\epsfxsize=8.0cm\epsfbox{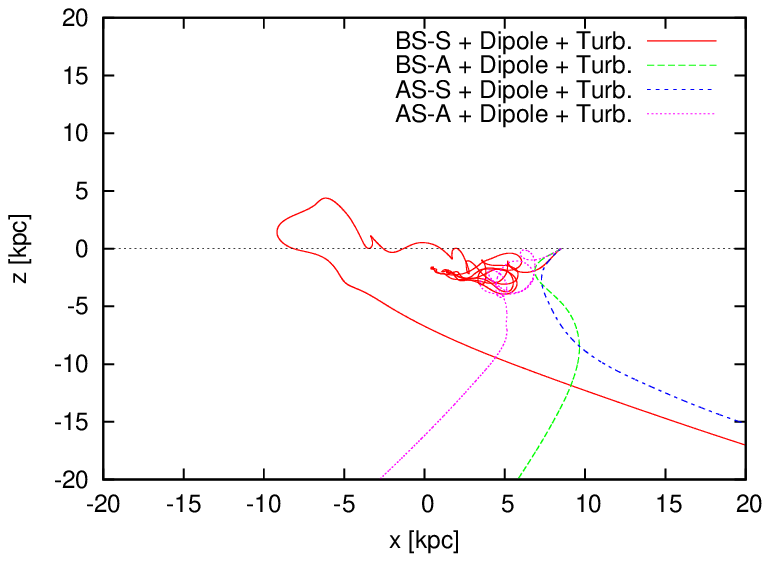}}}}
\figcaption{Examples of the trajectories of irons with the energy of $10^{19.8}$ eV projected onto a plane perpendicular to the Galactic disk and including the Galactic center and the Earth. The position of the Earth is $x = 8.5$ kpc. Their arrival directions are $(\ell, b) = (180.0^{\circ}, 0.3^{\circ})$. The BS-S model ({\it red}), the BS-A model ({\it green}), the AS-S model ({\it blue}), the AS-A model ({\it magenta}), are considered as a spiral component of GMF. The dipole component and the turbulent component are also taken into account.\label{fig:traFe}}
\vspace{0.3cm}

\begin{figure*}
\epsscale{0.85}
\plotone{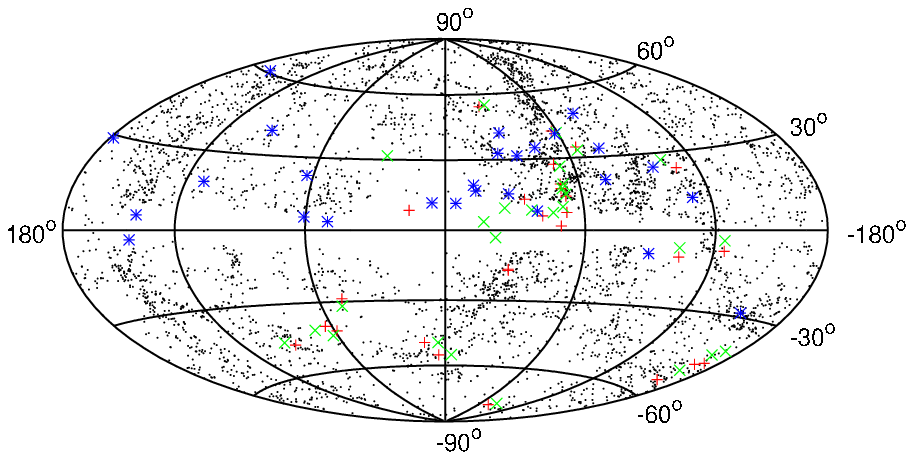}
\plotone{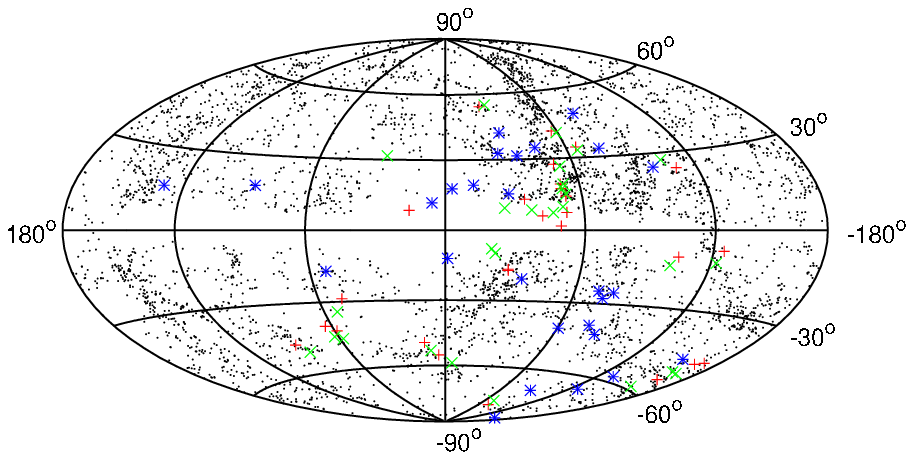}
\plotone{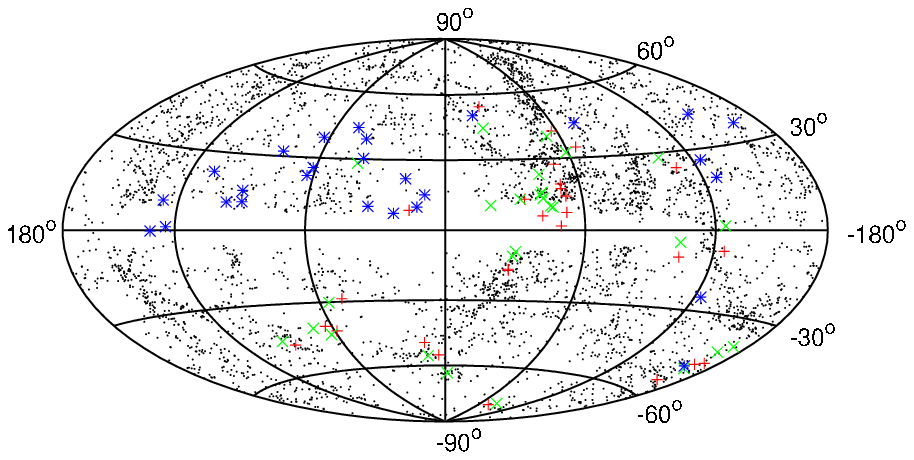}
\plotone{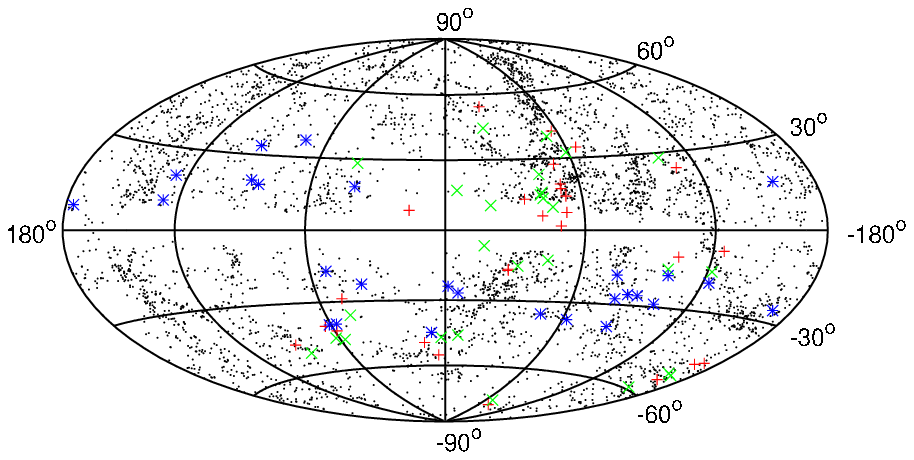}
\caption{Arrival directions of the PAO events before GMF deflections. The BS-S model ({\it upper left}), BS-A model ({\it upper right}), AS-S model ({\it lower left}), and AS-A model ({\it lower right}) are considered as a spiral component of GMF. The dipole component and the turbulent component of GMF are also taken into account. The composition of the PAO events is assumed to be protons ({\it green}) and irons ({\it blue}). The arrival directions of the PAO events are also plotted ({\it red}). Black points are IRAS galaxies within 75 Mpc.}
\label{fig:nuclei}
\end{figure*}

Fig. \ref{fig:nuclei} demonstrates the arrival directions of the PAO events before GMF deflections, assuming that all the PAO events are protons ({\it green}) or irons ({\it blue}). The BS-S model ({\it upper left}), the BS-A model ({\it upper right}), the AS-S model ({\it lower left}), the AS-A model ({\it lower right}) are considered as a spiral component of GMF. The dipole component and the turbulent component of GMF are also taken into account. These directions correspond to the positions of their sources if the effect of IGMF can be neglected. The detected arrival directions of the PAO events ({\it red}) and IRAS galaxies within $z = 0.018$ ({\it black}) are also plotted. In order to make this figure, the detected PAO events are backtracked in the 4 GMF models and their velocity directions just outside the Galaxy (at 40 kpc from the Galactic center) are plotted. The deflections of irons are quite large, and the correlation between the arrival directions (before GMF deflections) and the galaxy distribution is obviously lost. It is unnatural that the deflections by IGMF improve the correlation. Thus, a pure iron scenario is problematic in the viewpoint of the spatial correlation. Since light nuclei including protons may be contained in detected HECRs in a real situation, the situations can become better, but this figure points out that the contribution of GMF to the total deflections of nuclei are not negligible. In this case, the fraction of protons (or light nuclei) is an important parameter. This parameter also gives a hint of cosmic-ray composition at sources, i.e., a hint of the generation mechanism of HECRs. The arrival directions of irons before GMF modification are dependent on GMF modelling, as we can see in Fig. \ref{fig:nuclei}. Detailed modelling of GMF could also improve the correlation between the arrival directions of HECRs before GMF modification and their sources.

\subsubsection*{Acknowledgements:} 

H.T. thanks to Veniamin Berezinsky for useful comments at the early stage 
of this study and Peter Biermann for comments on GMF in the Galactic halo. 
This work is supported by World Premier International Research Center 
Initiative (WPI Initiative), the Ministry of Education, Culture, 
Sports, Science and Technology (MEXT) of Japan. 
This work is also supported by Grants-in-Aid for Scientific 
Research from the MEXT of Japan through No.21840019 (H.T.) 
and No.19104006 (K.S.).

\end{document}